\documentclass[11pt,a4paper]{article}
\pdfoutput=1

\usepackage{enumitem}
\usepackage{etex}
\usepackage{amsthm}
\usepackage{geometry}
\usepackage{dcolumn}
\usepackage{epsf}
\usepackage{mathrsfs}
\usepackage{multirow}
\usepackage{booktabs}
\usepackage{tabularx}
\usepackage{array}
\usepackage{slashed}
\usepackage{float}
\usepackage{leftidx}
\usepackage{setspace}
\usepackage{verbatim}
\usepackage{adjustbox}
\usepackage{tikz}
\usepackage[caption=false]{subfig}
\usepackage{arydshln}
\usepackage{jheppub}

\numberwithin{equation}{section}
\usetikzlibrary{arrows,decorations.markings,shapes.arrows,patterns,positioning}

\newcommand{\bea}{\begin{eqnarray}}
\newcommand{\eea}{\end{eqnarray}}
\newcommand{\bean}{\begin{eqnarray*}}
\newcommand{\eean}{\end{eqnarray*}}
\newcommand{\nn}{\nonumber\\}
\newcommand{\Sl}{\sum\limits}

\def\W #1{\widetilde{#1}}

\def\Label#1{\label{#1}%
  \smash{\hbox to0pt{\raise1ex\hbox{\tiny[#1]}\hss}}}
\def\Label#1{\label{#1}}
\renewcommand{\eqref}[1]{eq.~(\ref{#1})}

\newcommand{\figref}[1]{figure~\ref{#1}}

\newcommand{\secref}[1]{section~\ref{#1}}

\def\Tr{\mathop{\rm Tr}}

\def\Sl{\sum\limits}

\newcommand{\ctobedelete}[1]{}

\allowdisplaybreaks
\hypersetup{pageanchor=false}
\title{Explicit BCJ numerators of nonlinear sigma model}

\author{Yi-Jian Du ${}^{a}$, Chih-Hao Fu ${}^{b}$}

\affiliation[a]{Center for Theoretical Physics, School of Physics and Technology, Wuhan University \\ 299 Bayi Road, Wuhan 430072, China}
 \affiliation[b]{School of Mathematical Sciences, University of Nottingham \\University Park, Nottingham, NG7 2RD, UK  }
\emailAdd{yijian.du@whu.edu.cn }
 \emailAdd{chih.fu@nottingham.ac.uk}

\abstract{
In this paper, we investigate the color-kinematics duality in nonlinear sigma model (NLSM). We present explicit  polynomial expressions for the kinematic numerators (BCJ numerators). 
The calculation is done separately in two parametrization schemes of the theory using
Kawai-Lewellen-Tye relation inspired technique, both lead to polynomial numerators.
We summarize the calculation in each case  into a set of rules that generates  BCJ numerators
for all multilplicities.  In Cayley parametrization we find the numerator is described by a particularly simple formula solely
in terms of momentum kernel.
 }
\begin{document}
\maketitle
\hypersetup{pageanchor=true}

\section{Introduction}

The Color-Kinematics duality proposed by Bern, Carrasco and Johansson
(BCJ) in \cite{Bern:2008qj} suggests that the color factors and   momentum dependent numerators
of  Yang-Mills amplitudes, originally prescribed through Feynman
rules, can be reformulated using a set of graphical organization principles
that treats them on equal footing. The BCJ numerators associated
with each trivalent graph satisfies anti-symmetry and Jacobi identities
that mirrors the algebraic behavior of its color counterpart. It has
been realized that for such graphical organization to work, the amplitude
has to satisfy linear relations, and the revere is also true. Given
BCJ  amplitude relations, one can be assured of the existence of BCJ numerators
\cite{Monteiro:2011pc}. These amplitude relations have been proven both
from field theory \cite{Feng:2010my, Cachazo:2012uq} and string perspectives \cite{Stieberger:2009hq, BjerrumBohr:2009rd}.
The BCJ duality has a further independent, and yet perhaps even more
striking implication on gravity. When the color dependence of a Yang-Mills
amplitude is swapped in exchange  for  another copy of kinematic numerator,
the resulting double-copy formula reproduces gravity amplitude \cite{Bern:2008qj, Bern:2010ue}.
At tree level this prediction has been proven using multiple (Britto-Cachazo-Feng-Witten) BCFW
shifting \cite{Bern:2010yg}, and is understood to be equivalent
to the famous (Kawai-Lewellen-Tye) KLT relation \cite{KLT, Kiermaier}. At the moment of writing the
loop level correspondence remains a conjecture, however it has been
verified through accumlating evidence
\cite{Bern:2010ue, Bern:2012uf, Bern:2013qca, Bern:2014sna, Mafra:2014oia, Mafra:2014gja, Mafra:2015mja,
Boels:2013bi, Carrasco:2011mn, Bjerrum-Bohr:2013iza, Carrasco:2012ca, Bern:2013yya, Nohle:2013bfa,
Ochirov:2013xba, Chiodaroli:2013upa, Chiodaroli:2014xia, Yuan:2012rg, Oxburgh:2012zr, Saotome:2012vy}.
In addition, Color-Kinematics duality is also known to serve as one
of the criteria for counterterms and provides a guideline for the
UV behavior of gravity theory. For more details we refer the readers
to the comprehensive reviews \cite{Elvang:2013cua,Carrasco:2015iwa} and the references within.

In viewing of this apparent symmetry between the color and kinematics,
it is  tempting to think that a simple algebraic explanation might be
responsible for the behavior of BCJ numerators,  similar to that
of its color counterpart. And indeed, a partial understanding has
been achieved by studying the self-dual sector, where the cubic vertex
is identified as the structure constant of area-preserving diffeomorphism
algebra \cite{BjerrumBohr:2012mg,Monteiro:2011pc}. This explains 
at tree level the all-except-one
plus helicity in $D>4$ dimensions\footnote{ 
In four dimensions this amplitude is trivial. However it was noticed that the diffeomorphism algebra explains
the diagrammatically related all plus helicity integrand at one loop level, and in addition 
the one leg off-shell continued all-except-one plus helicity current \cite{BjerrumBohr:2012mg,Monteiro:2011pc}. 
}
and the ( maximally-helicity-violating) MHV amplitudes to all multiplicities. Beyond MHV, not very much is understood about the origin of
its algebraic behavior,\footnote{Note however, that the self-dual based understanding covers all helicity
configurations if one choses to work in the framework of scattering
equations \cite{Cachazo:2013gna, Cachazo:2013hca, Cachazo:2013iea}, where
the effective cubic vertices are parametrized
by nonlocal solutions \cite{Monteiro:2013rya}. Another notable
recent progress shows that the self-dual language can be actually
generalized one step further to the MHV amplitudes at one loop level,
by taking the infinite tension limit of dimensionally reduced string
amplitude \cite{He:2015wgf}.} partly because determining BCJ numerators
can be techinically
challenging. The complexity involves in the perturbative calculation
grows factorially as mutiplicity increses. In addition the numerators
are known to be non-unique. The presumably existing algebraic structure
can be easily obscured by generalized gauge degrees of freedom.

As it happens, Color-Kinematics duality is known to be respected by
a number of different theories not limited to Yang-Mills \cite{Johansson:2015oia, Bargheer:2012gv, Huang:2012wr, Monteiro:2014cda,
Luna:2015paa, Johansson:2014zca, Chiodaroli:2013upa, Sondergaard:2009za,
Weinzierl:2014ava, Naculich:2014naa, Naculich:2015zha}. It has already been pointed to exist also in e.g., QCD \cite{Chiodaroli:2015rdg}, Spontaneously Broken Einstein-Yang-Mills supergravity \cite{Johansson:2015oia}.
The focus of this paper is on the nonlinear sigma model (NLSM), described
by the chiral Lagrangian  originally designed to capture the phenomenological
behavior of the Golstone bosons corresponding to the
 $G\times G\rightarrow G$ isospin symmetry breaking. In
this model the flavor group $SU(N_{f})$ plays a similar role to the
color in strong  interactions and can be used to define partial amplitudes
and flavor ordered Feynman rules \cite{Kampf:2013vha}. (For this reason we
shall be using the terms color and flavor interchangeably in this paper.)
It was shown earlier that amplitudes of the nonlinear sigma model
(NLSM) satisfy BCJ relation   which has an off-shell extension in Cayley parametrization \cite{Chen:2013fya, Chen:2014dfa}. This relation actually implies the color-kinematic duality, as in Yang-Mills theory.

We feel that the NLSM serves as an interesting  practical setting to hopefully a better
understanding of the kinematic algebra, in the sense that unlike the
bi-adjoint scalar theory, the algebraic property is not a built-in
feature of the theory, and yet the theory is known to satisfy Color-Kinematics
duality, therefore leaving a puzzle as to identifying the responsible
algebra, or whether there is one existing, which is similar to the puzzle posed by
Yang-Mills theory. It is also not by construction a cubic theory and
can perhaps therefore provide us an example as to how to systematically
cope with contact terms. In addition, the duality is known to be respected
by the NLSM in arbitrary dimensions, which is still another feature
shared with Yang-Mills amplitudes. The NLSM numerator however, stands
a good chance to be formally simpler than those of Yang-Mills because it is given
by a scalar field theory and the corresponding Feynman rules completely
devoid of polariztion vectors, so that they can only be composed of
Mandelstam  variables.

In this paper we use the KLT inspired prescription to calculate NLSM
numerators \cite{Kiermaier}. We choose to work in the setting when one
particular leg is taken off-shell, allowing the propagator matrix
to be inverted so that the numerators can be reversely determined
in terms of amplitudes. The BCJ relations observed in \cite{Chen:2013fya}
formulated in Cayley parametrization scheme are repeatedly used to
cancel poles, much like the procedure introduced in \cite{Du:2011js}
to prove the color-dressed KLT relations, until an explicit expression
is obtained. The resulting numerator is completely pole-free
and
 is expressible as a sum of momentum kernel
\bea
n_{1|2,3,\dots,n-1|n}=\Sl_{\sigma}{\cal S}[n-1 \dots 2\mid\sigma]
\eea
over a subset of permutations explained in section \ref{sec:Graphs}.
In Cayley parametrization
the numerators derived from this procedure will pick up a special
set of basis numerators because of the asymmetric analytic continuation.
As we will explain more at the beginning of  section \ref{sec: pion-numerators}
 a generic numerator in this
picture is by construction defined through basis numerators, which
in turn are built from amplitudes. The numerators themselves do not
have to  possess symmetry with respect to permuting external lines.
As an alternative, we present the NLSM numerator in a more symmetrical
setting. We obtain BCJ relations following similar derivations to
\cite{Chen:2013fya} and compute the permutation symmetric numerators up to
$8$ points.

This paper is organized as follows. In section \ref{sec:NLSM} we briefly review
the BCJ relations between NLSM amplitudes in Cayley parametrization
and the KLT inspired prescription for  BCJ numerators.
 In \secref{sec:GeneralRule}, we present and prove general rules for
 constructing numerators in Cayley parametrization. Section \ref{sec:Graphs} provides
 a graphical summary for these rules. Permutation symmetric numerators
 are presented in section \ref{sec: pion-numerators}.
We conclude this paper in section \ref{sec: conclusion}. Eight-point permutation symmetric numerators are presented in the appendix.

\section{Preliminaries: Color-kinematic duality and amplitude relations in NLSM}\label{sec:NLSM}

In this section we briefly review color-kinematics duality and amplitude relations in  nonlinear sigma model
necessary for the discussions in this paper.

\subsection{KLT relation, color-kinematics duality and dual color decomposition}

Ever since the discovery of the squaring identities between gravity and Yang-Mills amplitudes
of Kawai, Lewellen and Tye (KLT) \cite{KLT}, it has been proven that the relation applies to
a number of different theories as well, by taking the field theory limit of various closed and
open string amplitudes  (see, {\it e.g.} \cite{Bern:2002kj} and the references therein).
For example it was realized that the full Yang-Mills
amplitude $ M(1,2,\dots,n)$ factorizes into products of a color-dressed scalar amplitude  $\W A(n,\sigma_{2,n-1},1)$ and a color-ordered Yang-Mills
amplitude  $A(1,\rho_{2,n-1},n)$ \cite{Bern:1999bx}. These relations are known to be expressible in manifestly $(n-2)!$
permutation symmetric form

%
\bea
 { M}(1,2,\dots,n)=(-1)^n\Sl_{\sigma,\rho}\W A(n,\sigma_{2,n-1},1){{\cal S}[\sigma_{2,n-1}\mid\rho_{2,n-1}]\over k_n^2}A(1,\rho_{2,n-1},n)
\Label{eq:n2-symm-klt}
\eea
%
and the $(n-3)!$ symmetric form
\bea
 M(1,2,\dots,n)=(-1)^{n+1}\Sl_{\sigma,\rho}\W A(n-1,n,\sigma_{2,n-1},1){\cal S}[\sigma_{2,n-2}|\rho_{2,n-2}]A(1,\rho_{2,n-2},n-1,n),
\Label{eq:n3-symm-klt}
\eea
 where $\sigma_{2,\dots,j}$ denotes permutations of $2,3,\dots, j$, and the momentum kernel is defined by
\bea
{\cal S}[i_1,\dots,i_k|j_1,\dots,j_k]=\prod\limits_{t=1}^k\left(s_{i_t1}+\Sl_{q>t}^k\theta(i_t,i_q)s_{i_ti_q}\right).\Label{eq:momentum-kernel}
\eea
 Both \eqref{eq:n2-symm-klt} and \eqref{eq:n3-symm-klt} treat the color scalar and the  color-ordered Yang-Mills  amplitudes equally.
Such symmetry has been made completely manifest by
the color-kinematics duality statement of
Bern Carrasco and Johannson \cite{Bern:2008qj},
which suggests a bi-cubic formulation of the full Yang-Mills amplitude
%
\begin{equation}
{M}(1,2,\dots,n)=\sum_{\text{cubic diags }  i}\frac{c_{i}n_{i}}{\prod_{\alpha_{i}}s_{\alpha_{i}}}.\Label{eq:cubic-exprssn}
\end{equation}
%
The kinematic 
numerators $n_i$ are assumed to satisfy the same algebraic identities  as their color counterparts  $c_i$
\bea
{\rm antisymmetry}:&~~& c_i\rightarrow-c_i\Rightarrow n_i\rightarrow-n_i\nn
{\rm Jacobi-like~identity}:&~~& c_i+c_j+c_k=0\Rightarrow n_i+n_j+n_k=0.~~\label{BCJ-duality}
\eea
%
Furthermore, it was realized that when the color factors $c_i$ are replaced by
another copy of the BCJ numerators, equation (\ref{eq:cubic-exprssn}) reproduces gravity \cite{Bern:2008qj, Bern:2010ue}.
In viewing of the fact that the same algebraic properties are shared between color and kinematic structures,
it is natural to expect that  various BCJ-dual color decompositions also describe Yang-Mills amplitude.
In particular it can be decomposed using  half ladder kinematic factors $n_{1|\sigma(2,..,n-1)|n}$,
%
\bea
M(1,2,\dots,n)=  \sum_{ \sigma} n_{1|\sigma_{2,n-2}|n}\W A(1,\sigma_{2,n-2},n),
\label{eq:ddm}
\eea
%
similarly to the decomposition illustrated in \cite{DelDuca:1999rs},
 which we shall refer to  as the dual Del Duca-Dixon-Maltoni (DDM) form in this paper.

\subsubsection{The construction of BCJ numertators from KLT relation}

In practice, solving the BCJ numerators can be a challenging task even if the BCJ duality
in a theory has been verified. Considering the fact that amplitudes depends on numerators only
linearly, in principle one can determine the numerators from amplitudes by inverting the propagator
matrix. However a subtlety arises because of the singular nature of the propagator matrix. It is known
that the inverse is not unique and one has to choose a specific generalized gauge
\cite{Boels:2012sy}. Notably one formally simple set of solutions was found using the KLT orthogonality
\cite{Cachazo:2013iea}. And indeed,  as a matter of fact it was realized earlier that 
one can readily arrive at a prescription
of the numerators in terms of momentum kernel
by comparing the (n-3)! symmetric form
of the KLT relation (\ref{eq:n3-symm-klt})
with the dual color decomposition formula (\ref{eq:ddm})
 \cite{Kiermaier, BjerrumBohr:2010hn, Naculich:2014rta, Fu:2014pya}.



Given the above set of solutions, because of the generalized gauge degrees of freedom, identifying
the algebraic structure can still be quite non-trivial. Generically the half ladder numerators obtained
through the algorithm just described may not simply correspond to a string of structure constants,
but to a gauge transformed, linear combination of several strings of structure constants. An alternative
solution to this dilemma is to tentatively make the propagator matrix non-singular through
analytic continuation, and take the on-shell limit afterwards. In the language of KLT this corresponds
to comparing the   $(n-2)!$ symmetric form (\ref{eq:n2-symm-klt})  with the dual color decomposition,
and write\footnote{Notice that the color scalar amplitude has reflection symmetry $\W A(1,\sigma_{2,n-1},n)=(-1)^n\W A(n,\sigma^T_{2,n-1},1)$.}
\bea
n_{1|\alpha_{2,n-2}|n}=\Sl_{\beta}{{\cal S}[\alpha_{2,n-2}^T|\beta_{2,n-2}]\over k_n^2}A\left(1,\beta_{2,n-2},n\right).
\Label{eq:n2-klt-prescription}
\eea
It has been recently proven by Mafra in \cite{Mafra:2016ltu} that such prescription agrees with the
Berends-Giele construction in the case of a bi-adjoint scalar theory, therefore yielding the non-gauge
transformed algebraic factor as the half ladder numerator.

At first sight, equation (\ref{eq:n2-klt-prescription}) may not be very ``good-looking''
because the vanishing  $k_n^2$
seems to lead to  a divergence.
However we will see in the following that this  divergence is canceled by BCJ relations.
In section \ref{sec:Cayley} we will start from this $(n-2)!$ permutation symmetric form
of the numerator prescription and see that the right hand side of this equation
actually completely reduced to polynomials of  Mandelstam variables  $s_{ij}$.
The resulting BCJ numerators therefore carries no pole.


\subsection{Cayley parametrization and the nonlinear sigma model}

In this paper we are interested in finding the BCJ numerators of the $SU(N_{f})$
nonlinear sigma model. This is the effective model describes the low energy behavior
of Goldstone bosons associated with $SU(N_{f})\times SU(N_{f}) \rightarrow SU(N_{f})$
symmetry breaking. The tree level amplitude of the Goldstone bosons was previously
observed to  possess color-kinematics duality \cite{Chen:2013fya}.


\subsubsection{Feynman rules  and Berends-Giele currents in Cayley parameterization}

The Lagrangian of the $SU(N_{f})$ non-linear sigma model is
\bea
\mathcal{L}={F^2\over 4}\Tr (\partial_{\mu}U\partial^{\mu}U^{\dagger}),
\eea
where $F$ is the decay constant. Each $U$ is defined in 
Caylay parametrization as
%
\bea
U=1+2\Sl_{n=1}^{\infty}\left({1\over 2F}\phi\right)^n,~~~~\label{Cayley}
\eea
where $\phi=\sqrt{2}\phi^at^a$, and $t^a$ are generators of the $SU(N_{f})$ flavor Lie algebra.
{It} was demonstrated in \cite{Kampf:2012fn,Kampf:2013vha} that partial amplitudes and
flavor-ordered Feynman rules can be defined by complete analogy to the color-ordering of Yang-Mills.
For this reason we will abuse the terminology a bit in this paper and 
 simply refer 
the $SU(N_{f})$ flavor as color. 
For future references we list the flavor ordered vertices as follows.
%


%
\bea
V_{2n+1} & = & 0, \nn
V_{2n+2} & = & \left(-{1\over 2F^2}\right)^n\left(\Sl_{i=0}^np_{2i+1}\right)^2=\left(-{1\over 2F^2}\right)^n\left(\Sl_{i=0}^np_{2i+2}\right)^2.\Label{Feyn-rules}
\eea
%
It is known that vertices of odd multiplicities vanish in the Cayley parametrization scheme.  Thus when  an
 $n$-point amplitude of NLSM is referred, unless otherwise mentioned the $n$ is always understood as an even number.
Note that the two expressions on the right hand side of equation (\ref{Feyn-rules})
are equivalent when momentum conservation is taken into account.

 Given the Feynman rules above, we can construct tree level off-shell currents  with one off-shell line through Berends-Giele recursion
\bea
&&J(1,2,...,n-1)\nn
&=&\frac{i}{P_{1,n-1}^2}\Sl_{m=4}^n\Sl_{0=j_0<j_1<\cdots<j_{m-1}=n-1}i V_{m}(P_{j_0+1,j_1},\cdots,P_{j_{m-2}+1,n-1},p_n=-P_{1,n-1})\times\prod\limits_{k=0}^{m-2} J(j_k+1,\cdots,j_{k+1}),\Label{B-G}\nn
\eea
where $p_n=-P_{1,n-1}\equiv-(p_1+p_2+\dots+p_{n-1})$. The starting point of this recursion is $J(1)=J(2)=\dots=J(n-1)=1$. 
 Note that building up an odd multiplicity current (including the off-shell line) requires
at least one odd multiplicity vertex, 
therefore  $J(2,\dots,2m+1)=0$.
Even multiplicity currents in general are bulit up only by odd numbers of even sub-currents
(i. e., those sub-currents with an odd number of on-shell lines and one off-shell line).


\subsubsection{Amplitude relations in NLSM}

It was pointed out in \cite{Chen:2013fya} that off-shell currents in Cayley parametrization satisfy 
an off-shell version of the $U(1)$ decoupling identity and the fundamental BCJ relation,
both of these two relations are restored to their more familiar original forms
in the on-shell limit.   
Furthermore, it was noted that the $(n-2)!$ symmetric formula of KLT relation is also 
restored  on-shell. 
Explicit formulas of these relations are given as follows.
\begin{itemize}
\item {\bf\emph{$U(1)$ identity in NLSM}}

The $U(1)$ identity for off-shell currents is given by
\bea
\Sl_{\sigma\in OP(\{\alpha_1\}\bigcup\{\beta_1,\dots,\beta_{2m}\})}J(\{\sigma\})={1\over 2F^2}\Sl_{divisions\{\beta\}\rightarrow\{B_{1}\},\{B_{2}\}}J(\{B_{1}\})J(\{B_{2}\}),\Label{off-shell-U(1)}
\eea
where, on the left hand side, we sum over all the possible permutations with  the relative order in set $\{\beta\}$  and 
 the relative order in set $\{\alpha\}$ kept fixed. In this paper we only need to consider when
there is only one element $\alpha_1$ in $\{\alpha\}$ (the $U(1)$ decoupling identity) although generically the number
 of elements in $\{\alpha\}$ and $\{\beta\}$ can be quite arbitrary.
 On the right hand side, we divide the ordered set $\{\beta_1,\dots,\beta_{2m}\}$ into two nonempty ordered subsets,
 each containing an odd number of elements. 
 For example, suppose if we have six $\beta$'s in the original set $\{\beta\}$,
 the following three different distributions of $\beta$'s into $\{B_{1}\}$ and $\{B_{2}\}$ should be included in the summation:
  $\{B_1\}=\{\beta_1\}$, $\{B_{2}\}=\{\beta_2,\dots,\beta_6\}$; 
  $\{B_1\}=\{\beta_1,\beta_2,\beta_3\}$, $\{B_{2}\}=\{\beta_4,\beta_5,\beta_6\}$ 
  and $\{B_1\}=\{\beta_1,\dots,\beta_5\}$, $\{B_{2}\}=\{\beta_6\}$.

\item {\bf\emph{Fundamental BCJ relation in NLSM}}

The fundamental BCJ relation is  expressible  as
%
\bea
&&\Sl_{\sigma\in OP(\{\alpha_1\}\bigcup\{\beta_1,\dots,\beta_{2m-1}\})}  
\left( 
\Sl_{\xi_{\sigma_i}<\xi_{\alpha_1}}
 s_{\alpha_1\sigma_i}+s_{\alpha_1n}\right)J(\{\sigma\},\beta_{2m})\nn
&=&-{1\over 2F^2}\Sl_{divisions\{\beta\}\rightarrow\{B_{1}\},\{B_{2}\}}\left[\Sl_{\beta_i\in\{B_{2}\}}s_{\alpha_1\beta_i}J(\{B_{1}\})J(\{B_{2}\})\right],\Label{off-shell-BCJ0}
\eea
%
where we use $\xi_i$ to denote the position of the leg $i$ in permutation $\sigma$ and   we always have a term  $s_{\alpha_1n}\equiv 2p_{\alpha_1}\cdot p_n$ in the coefficients of each currents on the left hand side\footnote{This notation is slightly different from that given in \cite{Chen:2013fya}, where the off-shell leg is denoted by $1$ and the factor $s_{\alpha_11}$ is hidden by setting $\xi_1=0$.}. On the right hand side,  the summation runs over all the possible  distributions  of the ordered set $\{\beta\}$ into two sub-ordered sets $\{B_{1}\}$ and $\{B_{2}\}$. Since $J(\{B_{1}\})$ or $J(\{B_{2}\})$ must vanish when $\{B_1\}$ or $\{B_{2}\}$ 
 is even, the surviving distributions 
 are those with both 
 $\{B_{1}\}$ and $\{B_{2}\}$
 odd.

When we multiply a factor $-s_{\alpha_1n}$ to both sides of the $U(1)$ identity \eqref{off-shell-U(1)} and add it to the fundamental BCJ relation \eqref{off-shell-BCJ}, we get another form of the fundamental BCJ relation 
%
\bea
&&\Sl_{\sigma\in OP(\{\alpha_1\}\bigcup\{\beta_2,\dots,\beta_{2m}\})}\Sl_{\xi_{\sigma_i}<\xi_{\alpha_1}}s_{\alpha_1\sigma_i}J(\beta_1,\{\sigma\})\nn
&=&{1\over 2F^2}\Sl_{divisions\{\beta\}\rightarrow\{B_{1}\},\{B_{2}\}}\left[\Sl_{\beta_i\in\{B_{1}\}}s_{\alpha_1\beta_i}J(\{B_{1}\})J(\{B_{2}\})\right],\Label{off-shell-BCJ}
\eea
%
(Assuming momentum conservation $-\left(\Sl_{\beta_i\in\{B_2\}}p_{\beta_i}+p_n\right)=\left(\Sl_{\beta_i\in\{B_1\}}p_{\beta_i}+p_{\alpha_1}\right)$ and on-shell condition $p_{\alpha_1}^2=0$ 
are holding.) 
We will be using this alternative form of the fundamental BCJ relation frequently in the following sections.

\item {\bf\emph{KLT relations for color-dressed amplitudes in NLSM}}

The on-shell color-dressed amplitudes of nonlinear sigma model are also known to satisfy the same KLT relation as  
Yang-Mills amplitudes do.
%
 Thus we have{\footnote{  Note that the $(-1)^n$ factor in the standard $(n-2)!$ symmetric KLT formula have been dropped because $n$ is always an even number. }
\bea
M(1,2,\dots,n)=\Sl_{\sigma,\rho }\W A(n,\sigma_{2,n-2},1){{\cal S}[\sigma_{2,n-2}\mid\rho_{2,n}]\over k_n^2} A(1,\rho_{1,2},n).
\eea}
In the next section, we will start from this $(n-2)!$ formula to construct BCJ numerators.
\end{itemize}

\section{The rule for BCJ numerator in NLSM with Cayley parametrization}
\label{sec:Cayley}
As already stated in section 2, once we have all BCJ numerators in dual-DDM formula, we can always construct other numerators  using  antisymmetry and Jacobi identities.
The BCJ numerator in dual DDM formula can be directly read off from the $(n-2)!$  symmetric formula of KLT relation, although this formula contains a regulator $k_n^2\to 0$ in the denominator.
 In this section, we present a set of systematic construction rules of the NLSM numerator using the $(n-2)!$ symmetric formula.
 We find that  starting from extending the momentum of  leg $n$ off  its  mass shell, we can reduce the numerator, which
  was originally  expressed  as  \eqref{eq:n2-klt-prescription},
into  simpler form containing
only polynomials of  Mandelstam variables, with no pole or color-ordered amplitudes in the final  expression. 
Thus the limit $k_n^2\to 0$ can be taken directly.
The main idea is the following.
\begin{itemize}

\item We define the off-shell extension $N_{1|\sigma_{2,n-1}|n}$ of BCJ numerator $n_{1|\sigma_{2,n-1}|n}$ (see \eqref{eq:n2-klt-prescription}) as
\bea
 N_{1|\sigma_{2,n-1}|n}=\Sl_{\rho}{{\cal S} [\sigma_{2,n-1}^T\mid\rho_{2,n-1} ]} J(1,\rho_{2,n-1}),\Label{eq:Off-shell-numerator}
\eea
where  the $J(1,\rho_{2,n-1})$ are currents with leg $n$  taken off-shell    defined recursively through Berends-Giele recursion relation (\ref{B-G}). When  leg $n$ goes on-shell, $k_n^2J(1,\rho,n)\to A(1,\rho,n)$, therefore this off-shell extension  returns to
 the
on-shell expression \eqref{eq:n2-klt-prescription} of $n(1,\sigma,n)$   read off from  the $(n-2)!$  symmetric  KLT relation.
\item Applying the off-shell $U(1)$ identity and fundamental BCJ relations  repeatedly, we reduce $N_{1|\sigma_{2,n-1}|n}$ into polynomials of $s_{ij}$. 
 The final expression of $n_{1|\sigma_{2,n-1}|n}$
is then obtained by taking the on-shell limit.

\end{itemize}

\subsection{Six-point example}\label{SixPointEg}
Before giving the general rule, let us begin with a warm-up example : The numerators  of tree-level six-point NLSM amplitudes. 
We consider the six-point numerator in dual DDM decomposition  $n_{1|2345|6}$. Numerators of the form  $n_{1|\sigma_{2,n-1}|6}$ can be obtained by a relabeling.
Other numerators are produced by Jacobi identity and antisymmetry.

%
From definition \eqref{eq:Off-shell-numerator}, the off-shell extension of $n_{1|2345|6}$ is given by
\bea
 N_{1|2345|6}&=&\Sl_{\rho}{\cal S}[5432|\rho_{2,5}]J\left(1,\rho_{2,5}\right).\Label{6ptNumerator}
\eea
 To express this  {$N_{1|2345|6}$  as  polynomial of Mandelstam variables}, we  simplify the  right hand side of \eqref{6ptNumerator} level by level as follows.

\begin{itemize}
\item {\bf Step-1} We first rewrite the sum over permutations of $\{2,3,4,5\}$ as $ \Sl_{\gamma}\Sl_{\alpha\in\text{OP}(5\cup\{\gamma_{2,4}\})}$, thus the off-shell extension of numerator   $n_{1|2345|6}$ becomes
 \bea
N_{1|2345|6}&=&\Sl_{\gamma}\Sl_{\alpha\in\text{OP}(5\cup\{\gamma_{2,4}\})}{\cal S}[5432|\alpha]J\left(1,\{\alpha\}\right).\Label{Eq:6pt-1}
\eea
%
From the definition of momentum kernel (\ref{eq:momentum-kernel}), we have
\bea
{\cal S}[5432|\gamma_2\dots\gamma_i5\gamma_{i+1}\dots\gamma_4]=s_{5\gamma_j}{\cal S}[432|\gamma],
\eea
where $\{\gamma_2,\gamma_3,\gamma_4\}$ can be any given permutation of $2$, $3$, $4$. Then the  right hand side of \eqref{Eq:6pt-1} reads
\bea
N_{1|2345|6}&=&\Sl_{\gamma\in\text{perm}\{2,3,4\}}{\cal S}[432|\gamma]\Bigl[s_{51}J(1,5,\gamma_2,\gamma_3,\gamma_4)+(s_{51}+s_{52})J(1,\gamma_2,5,\gamma_3,\gamma_4)\\
&&+(s_{51}+s_{52}+s_{53})J(1,\gamma_2,\gamma_3,5,\gamma_4)+(s_{51}+s_{52}+s_{53}+s_{54})J(1,\gamma_2,\gamma_3,\gamma_4,5)\Bigr].\nonumber
\eea
The factor in brackets is just the left hand side of the 
 NLSM off-shell BCJ relation, 
thus   $N_{1|2345|6}$ can be 
 re-expressed as
 \bea
N_{1|2345|6}&=&\Sl_{\gamma\in\text{perm}\{2,3,4\}}(s_{51}+s_{5\gamma_L}){\cal S}[432|\gamma]\left[\Sl_{\text{Divisions}\gamma\to\gamma_L\gamma_R}J\left(1,\gamma_L\right)J\left(\gamma_R\right)\right] ,\Label{Eq:6pt-2}
\eea
 where in  the brackets the summation runs over 
all permutations such that each of   the $J\left(1,\gamma_L \right)$ and $J\left(\gamma_R\right)$  contains 
 an odd number of on-shell legs. 
 Specifically, for a 
given permutation $\{\gamma_2,\gamma_3,\gamma_4\}$, we have  the following two products in the sum over divisions
 \bea
J(1)J(\gamma_2,\gamma_3,\gamma_4),&~~~~~& J(1,\gamma_2,\gamma_3)J(\gamma_4).
\eea

\item{\bf Step-2}  After the simplification in step-1,  
the number of elements in the momentum kernel is reduced by one, and the six-point currents are reduced to two lower-point ones which do not contain  leg $5$.
In \eqref{Eq:6pt-2}, the sum over permutations  $\gamma\in\text{perm}\{2,3,4\}$ can be further  re-expressed as  $\Sl_{\xi}\Sl_{\text{OP}\left(4\cup \xi_{2,3}\right)}$,
 so that the momentum dependence on 
leg $4$ in each momentum kernel ${\cal S}[432|\gamma]$  factorizes 
\bea
{\cal S}\left[432|4\xi_2\xi_3\right]&=&s_{41}{\cal S}\left[32|\xi_2\xi_3\right],\nn
{\cal S}\left[432|\xi_24\xi_3\right]&=&\left(s_{21}+s_{4\xi_2}\right){\cal S}\left[32|\xi_2\xi_3\right],\nn
 {\cal S}\left[432|\xi_2\xi_34\right]&=&\left(s_{41}+s_{4\xi_2}+s_{4\xi_3}\right){\cal S}\left[32|\xi_2\xi_3\right].
\eea
Thus the two possible divisions in \eqref{Eq:6pt-2} can further be rearranged as follows.
\begin{itemize}
\item  The sum of terms containing currents of the form  $J(1)J(\gamma_2,\gamma_3,\gamma_4)$ is given by
\bea
&&s_{51}\Sl_{\xi\in \text{perm}\{23\}}{\cal S}\left[32|\xi_2\xi_3\right]J(1)\Biggl[s_{41}J(4,\xi_2,\xi_3)+\left(s_{41}+s_{4\xi_2}\right)J(\xi_2,4,\xi_3)\nn
&&~~~~~~+\left(s_{41}+s_{4\xi_2}+s_{4\xi_3}\right)J(\xi_2,\xi_3,4)\Biggr]=s_{51}\Sl_{\xi\in \text{perm}\{23\}}{\cal S}\left[32|\xi_2\xi_3\right]\left(s_{41}+s_{4\xi_2}\right).
\eea
Applying the off-shell $U(1)$ identity (\ref{off-shell-U(1)}) and fundamental BCJ relation (\ref{off-shell-BCJ}) on the sum in brackets and 
citing the explicit 
expression of  ${\cal S}\left[32|\xi_2\xi_3\right]$, we then write the above 
 solely in terms of  Mandelstam variables  as 
\bea
s_{51}\left[\left(s_{41}+s_{43}\right)s_{31}s_{21}+\left(s_{41}+s_{42}\right)(s_{31}+s_{32})s_{21}\right].\Label{Eq:6pt-3}
\eea

\item  The sum of terms containing currents of the form  $ J(1,\gamma_2,\gamma_3)J(\gamma_4)$ is given by
\bea
&&\Sl_{\xi\in\text{perm}\{2,3\}}(s_{51}+s_{54}+s_{5\xi_2}){\cal S}\left[32|\xi_2\xi_3\right]\Bigl[s_{41}J(1,4,\xi_2)+(s_{41}+s_{4\xi_2})J(1,\xi_2,4)\Bigr]\nn
&+&(s_{51}+s_{52}+s_{53})(s_{41}+s_{42}+s_{43})\Sl_{\xi\in\text{perm}\{2,3\}}{\cal S}\left[32|\xi_2\xi_3\right]J(1,\xi_2,\xi_3).
\eea
When applying the BCJ relation on the brackets of the first line and the second sum we obtain
\bea
&&(s_{51}+s_{54}+s_{53})s_{41}s_{31}s_{21}+(s_{51}+s_{54}+s_{52})s_{41}(s_{31}+s_{32})s_{21}\nn
&+&(s_{51}+s_{52}+s_{53})(s_{41}+s_{42}+s_{43})s_{31}s_{21}. \Label{Eq:6pt-4}
\eea

\end{itemize}

\item{\bf Step-3} Finally, putting \eqref{Eq:6pt-3} and \eqref{Eq:6pt-4} together, we get the expression for numerator 
%
\bea
 N_{1|2345|6}&=&s_{51}\left[\left(s_{41}+s_{43}\right)s_{31}s_{21}+\left(s_{41}+s_{42}\right)(s_{31}+s_{32})s_{21}\right]\Label{6pt-numerator}\nn
&&+(s_{51}+s_{54}+s_{53})s_{41}s_{31}s_{21}+(s_{51}+s_{54}+s_{52})s_{41}(s_{31}+s_{32})s_{21}\nn
&&+(s_{51}+s_{52}+s_{53})(s_{41}+s_{42}+s_{43})s_{31}s_{21}.
\eea
 The above expression gives precisely the BCJ numerator   $n_{1|2345|6}$ as  $k_n^2\to 0$. 

\end{itemize}

\subsection{The general rule}\label{sec:GeneralRule}
From the six-point example
 we see that 
the numerator defined  through dual DDM decomposition
can be expressed as  polynomial of Mandelstam variables $s_{ij}$ 
if the off-shell BCJ relation is applied repeatedly.
The final expression
 contains no pole. 
Now let us generalize  the six-point example to arbitrary  higher-point cases.

The off-shell extension of general numerator  $n_{1|2,\dots,n-1|n}$ in dual DDM decomposition reads
 \bea
N_{1|2,\dots,n-1|n}=\Sl_{\rho}{\cal S}[n-1,n-2,\dots,2|\rho]J\left(1,\rho_{2,n-1}\right).
\eea
For convenience, 
 let us introduce 
a new notation  $\mathcal{N}\left(1\{\gamma_1\}|\{\gamma_2\}|\dots|\{\gamma_P\}\right)$  to be called 
 a Level-${I}$ factor
\bea
&&\mathcal{N}\left(1\{\gamma_1\}|\{\gamma_2\}|\dots|\{\gamma_P\}\right)\Label{eq:Level-Ifactor}\\
&\equiv&\Sl_{\sigma_{J=1\dots I}\in\text{perm}\{\gamma_J\}}{\cal S}\left[(n-I)\dots 2\mid\sigma_1\sigma_2\dots\sigma_P\right]J\left(1,\sigma_1\right)J\left(\sigma_2\right)\dots J\left(\sigma_I\right),\nonumber
\eea
where we  distribute  $2,3,\dots (n-I)$ into $P\leq I$ non-ordered sets $\{\gamma_1\}$, $\{\gamma_2\}$,...,$\{\gamma_P\}$ such that each of $\{1,\gamma_1\}$, ..., $\{\gamma_P\}$  can only contain odd number of elements.
There are two  boundary cases
\begin{itemize}
\item Level-$1$ factor is nothing but the  off-shell extended numerator $N_{1|2,\dots,n-1|n}$
 \bea
N_{1|2,\dots,n-1|n}=\mathcal{N}\left(1\{2,3,\dots,n-1\}\right).
\eea
\item Level-$(n-2)$ factor is
\bea
 {\cal N}\left(1\mid 2\right)=s_{21}J(1)J(2)=s_{21}.
\eea
\end{itemize}
To reduce the numerator into polynomial of $s_{ij}$, we should reduce the level-$1$ factor to the level-$(n-2)$ factor. This procedure can be achieved iteratively.
Let us consider a given level-$I$ factor $\mathcal{N}\left(1\{\gamma_1\}|\dots|\{\gamma_K\}|\dots|\{\gamma_I\}\right)$, where the largest element $n-I$ is in the set $\{\gamma_K\}$. From definition (\ref{eq:Level-Ifactor})  we have
\bea
&&\mathcal{N}\left(1\{\gamma_1\}|\dots|\{\gamma_K\}|\dots|\{\gamma_I\}\right)\nn
&=&\Sl_{\sigma_{J|J\neq K}\in\text{perm}\{\gamma_J\}}\Sl_{\rho\in\text{perm}\left(\{\gamma_K\}/(n-I)\right)}\Sl_{j=0}^{m}\Biggl[{\cal S}\left[(n-I)\dots 2\mid\sigma_1\dots\sigma_{K-1}\rho_1\dots\rho_j(n-I)\rho_{j+1}\dots\rho_m\sigma_{K+1}\dots\sigma_J\right]\nn
&&~~~~~~~~~~~~~~~~~~~~\times J\left(1,\sigma_1\right)\dots J\left(\rho_1\dots\rho_j(n-I)\rho_{j+1}\dots\rho_m\right)\dots J\left(\sigma_I\right)\Biggr],
\eea
where we have rewritten the sum over permutations $\sigma_K\in\text{perm}\{\gamma_K\}$ by $\Sl_{\rho\in\text{perm}\left(\{\gamma_K\}/(n-I)\right)}\Sl_{j=0}^{m}$.
Using factorization property of momentum kernel, we  re-express the momentum kernel  in the above equation  as
\bea
\left[\Sl_{p=1}^{K-1}s_{(n-1)\sigma_p}+\Sl_{q=1}^js_{(n-1)\rho_q}\right]
 {\cal S}[(n-I-1) \dots 2\mid\sigma_1\dots\sigma_{K-1}\rho_1\dots\rho_j\rho_{j+1}\dots\rho_m\sigma_{K+1}\dots\sigma_J].
\eea
Thus the factor $ {\cal N}\left(1\{\gamma_1\}|\dots|\{\gamma_K\}|\dots|\{\gamma_I\}\right)$ can be written as
\bea
&&\mathcal{N}\left(1\{\gamma_1\}|\dots|\{\gamma_K\}|\dots|\{\gamma_I\}\right)\Label{Eq:gen-1}\\
&=&\Sl_{\sigma_{J|J\neq K}\in\text{perm}\{\gamma_J\}}J\left(1,\sigma_1\right)\dots J\left(\sigma_{K-1}\right)  J\left(\sigma_{K+1}\right)\dots J\left(\sigma_I\right)\nn
&&\times\Sl_{\rho\in\text{perm}\left(\{\gamma_K\}/(n-I)\right)} {\cal S}[(n-I-1) \dots 2\mid\sigma_1\dots\sigma_{K-1}\rho_1\dots\rho_j\rho_{j+1}\dots\rho_m\sigma_{K+1}\dots\sigma_J]\nn
&&\times\left[\Sl_{j=0}^{m}\left(\Sl_{p=1}^{K-1}s_{(n-1)\sigma_p}+\Sl_{q=1}^js_{(n-1)\rho_q}\right)J\left(\rho_1\dots\rho_j(n-I)\rho_{j+1}\dots\rho_m\right)\right]\nonumber.
\eea
According to the off-shell  $U(1)$ identity (\ref{off-shell-U(1)}) and the fundamental BCJ relations (\ref{off-shell-BCJ})\footnote{For convenience, we set the coupling constant ${1\over 2F^2}$ to 1.}, we can replace the last line by
\bea
\Sl_{\text{Divisions}\rho\to\rho_L\rho_R}\left(\Sl_{p=1}^Ks_{(n-1)\sigma_p}+s_{(n-1)\rho_L}\right)J(\rho_L)J(\rho_R).
\eea
Therefore, the last two lines of \eqref{Eq:gen-1}   together become
\bea
&&\Sl_{\rho\in\text{perm}\left(\{\gamma_K\}/(n-I)\right)}\Sl_{\text{Divisions}\rho\to\rho_L\rho_R} {\cal S}[(n-I-1) \dots 2\mid\sigma_1\dots\sigma_{K-1}\rho_1\dots\rho_j\rho_{j+1}\dots\rho_m\sigma_{K+1}\dots\sigma_J]\nn
&&\times\left[\left(\Sl_{p=1}^{K-1}s_{(n-1)\sigma_p}+s_{(n-1)\rho_L}\right)J(\rho_L)J(\rho_R) \right].
\eea
%
%
Rearranging the two sums above, we get
\bea
\Sl_{\rho\in\text{perm}\left(\{\gamma_K\}/(n-I)\right)}\Sl_{\text{Divisions}\rho\to\rho_L\rho_R} \to \Sl_{\left(\{\gamma_K\}/(n-I)\right)\to\gamma_L\gamma_R}\Sl_{\rho_L\in\text{perm}\{\gamma_L\}}\Sl_{\rho_R\in\text{perm}\{\gamma_R\}}
\eea
 and we finally rewrite the factor $\mathcal{N}\left(1\{\gamma_1\}|\dots|\{\gamma_K\}|\dots|\{\gamma_I\}\right)$ by the following combination of level-$(I+1)$ factors
\bea
&&\mathcal{N}\left(1\{\gamma_1\}|\dots|\{\gamma_K\}|\dots|\{\gamma_I\}\right)\Label{Eq:Rule}\\
&=&\Sl_{\left(\{\gamma_K\}/(n-I)\right)\to\gamma_L\gamma_R}\left(\Sl_{p=1}^Ks_{(n-1)\sigma_p}+s_{(n-1)\gamma_L}\right)\mathcal{N}\left(1\{\gamma_1\}|\dots|\{\gamma_L\}|\{\gamma_R\}|\dots|\{\gamma_I\}\right)\nonumber.
\eea

In general, we start from the level-1 factor, using the iterative relation above step by step till we do not have any momentum kernel and nontrivial currents in the expression. We finally get the polynomial expression of off-shell extension  $N_{1|2,\dots,n-1|n}$ of BCJ numerator $n_{1|2,\dots,n-1|n}$.

\subsubsection{Revisiting the Six-point example}
  As a demonstration, let us try calculating again  the explicit  six point off-shell extenion $N_{1|2345|6}$ of the numerator $n_{123456}$  using  the iterative rule \eqref{Eq:Rule}  we just derived.
\begin{itemize}
\item We start from $N_{1|2345|6}=\mathcal{N}(1\{2345\})$. After removing leg $5$, we have
\bea
\mathcal{N}(1\{2345\})&=&s_{51}\mathcal{N}(1|\{234\})+(s_{51}+s_{52}+s_{53})\mathcal{N}(1\{23\}|4)+(s_{51}+s_{52}+s_{54})\mathcal{N}(1\{24\}|3)\nn
&&+(s_{51}+s_{53}+s_{54})\mathcal{N}(1\{34\}|2).
\eea
\item Now we remove $4$ from $\mathcal{N}(1|\{234\})$, $\mathcal{N}(1\{23\}|4)$, $\mathcal{N}(1\{24\}|3)$ and $\mathcal{N}(1\{34\}|2)$. By doing so we get
\bea
&&\mathcal{N}(1\{2345\})\nn
&=&s_{51}(s_{41}+s_{42})\mathcal{N}(1|2|3)+s_{51}(s_{41}+s_{43})\mathcal{N}(1|3|2)+(s_{51}+s_{52}+s_{53})(s_{41}+s_{42}+s_{43})\mathcal{N}(1\{23\})\nn
&&+(s_{51}+s_{52}+s_{54})s_{41}\mathcal{N}(1|2|3)+(s_{51}+s_{53}+s_{54})s_{41}\mathcal{N}(1|3|2).
\eea
\item Then removing $3$ from $(1|2|3)$, $(1|3|2)$ and $(1\{23\})$ and we arrive at
\bea
\mathcal{N}(1\{2345\})&=&s_{51}(s_{41}+s_{42})(s_{31}+s_{32})\mathcal{N}(1|2)+s_{51}(s_{41}+s_{43})s_{31}\mathcal{N}(1|2)\nn
&&+(s_{51}+s_{52}+s_{53})(s_{41}+s_{42}+s_{43})s_{31}\mathcal{N}(1|2)+(s_{51}+s_{52}+s_{54})s_{41}(s_{31}+s_{32})\mathcal{N}(1|2)\nn
&&+(s_{51}+s_{53}+s_{54})s_{41}s_{31}\mathcal{N}(1|2).
\eea
\item Finally, removing $2$ from $(1|2)$ and we obtain the following result.
\bea
N_{1|2345|6}&=&\mathcal{N}(1\{2345\})\nn
&=&s_{51}(s_{41}+s_{42})(s_{31}+s_{32})s_{12}+s_{51}(s_{41}+s_{43})s_{31}s_{12}\nn
&&+(s_{51}+s_{52}+s_{53})(s_{41}+s_{42}+s_{43})s_{31}s_{21}+(s_{51}+s_{52}+s_{54})s_{41}(s_{31}+s_{32})s_{21}\nn
&&+(s_{51}+s_{53}+s_{54})s_{41}s_{31}s_{21},
\eea
which precisely agree with the six-point result (\ref{6pt-numerator}) given in the previous section.
\end{itemize}

\subsection{Graphical rules and explicit numerator formula in Cayley parametrization}\label{sec:Graphs}


\begin{figure}
\centering
\includegraphics[width=0.7\textwidth]{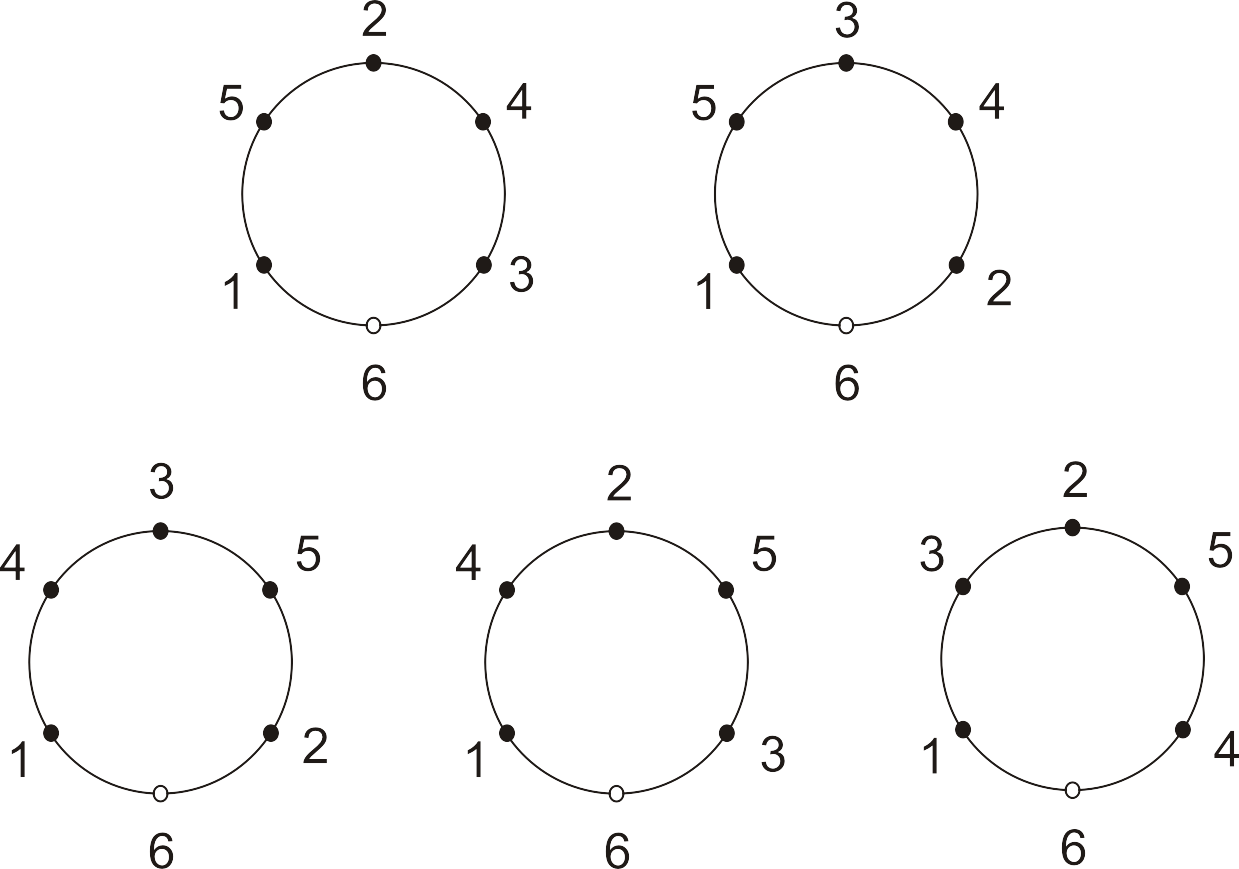}
\caption{All possible   configurations for the six-point numerator.}\label{Fig:6ptGraph1}
\end{figure}
\begin{figure}
\centering
\includegraphics[width=0.7\textwidth]{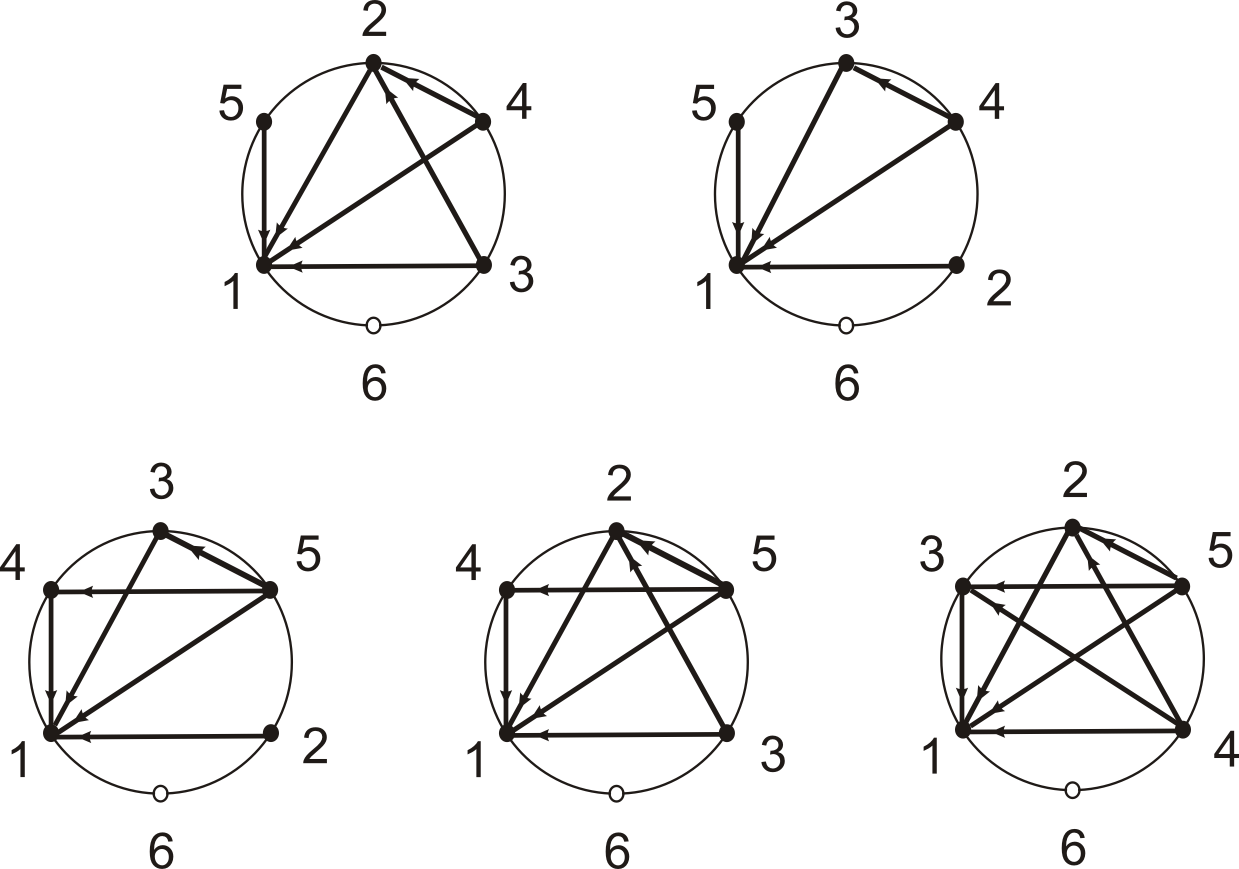}
\caption{All arrow graphs for the six-point numerator.}\label{Fig:6ptGraph2}
\end{figure}

The derivation just elaborated can be conveniently summarized by a set of conditions
used to determine the relative positions of external lines. For simplicity we choose to
represent these conditions as points on a circle. The explicit BCJ numerator will be
given by the sum of contributions read off from each circle once all the integer labels
have been assigned to these points.

%
\begin{itemize}
\item {\emph{Step-1}} To determine the $n$-point half ladder numerator $n_{1|2\dots n-1|n}$
in  Cayley parametrization, we begin  by  drawing $n$ points on a circle, with
one of them labeled by the integer $n$. We associate this special position with a hollow point
to emphasize that it is off-shell continued.
%
\item {\emph{Step-2}}  
The rest of the labels are then assigned in descending order: $n-1$, $n-2$, ..., $1$,  to the remaining points.
We request that when every label $j$ is assigned to a position,  either an odd number of unlabeled points 
or no unlabeled point is between  $j$ and the first labeled point on the left/right hand side of $j$.
When this assignment of all legs is completed, we further request  discarding
 all graphs where integers $1$ and $n$ are not adjacent. 
This
should give us a collection of graphs.
In the case of $6$ points, the collection of all configurations satisfying the above criteria
is shown in \figref{Fig:6ptGraph1}.
%
%
\item {\emph{Step-3}}  The graphs obtained through previous steps are subsequently
decorated with arrows.
For every labeled point $i \in \{i | 1<i<n \} $ we draw arrows from $ i $ to all of the labeled points  $ j < i $
that are on its left. Every such arrow is associated with a factor $s_{ij}$.
(A demonstration of the arrow graphs at $6$ points is given by  \figref{Fig:6ptGraph2})
%
\item {\emph{Step-4}}  Denoting the sum of associated factors with arrows
coming from point $i$ as $S_i=\Sl_{j<i,\xi_j<\xi_i}s_{ij}$, the
products $\prod_{j=2}^{n-1}S_j$ defines the contribution of each graph.
Summing over all graphs gives us the full BCJ numerator in Cayley parametrization.

%
%
\end{itemize}
%
Note that the conditions used to determine labels follows the operations in section \ref{sec:GeneralRule}
 in a more or less straight forward manner. Had we artificially inserted a leg $\alpha_{1}$
 between the two separated sets $\beta_{1}$ and $\beta_{2}$ on the right hand side of
 equations (\ref{off-shell-U(1)}), (\ref{off-shell-BCJ}) for the purpose of bookkeeping every time when a $U(1)$ identity or
 fundamental BCJ relation  was used to simplify the numerator formula (\ref{eq:Off-shell-numerator}),
 the resulting sequences of legs at the end of the calculation, when placed on a circle,
 will be the same as those in \figref{Fig:6ptGraph1}. The rest of the rules for constructing
 the numerator can be easily seen by keeping track of how the Mandelstam variables
 were assigned.


The outcome of this graphical construction can be expressed as a succinct formula in terms of momentum kernel
\bea
n_{1|2,3,\dots,n-1|n}=\Sl_{\sigma}{\cal S}[n-1 \dots 2\mid\sigma],
\label{eq:cayley-summarizing-result}
\eea
where $\sigma$ denote all possible permutations of $2$, $3$,...,$n-1$ satisfying the conditions in Step-2 and ${\cal S}$  is momentum kernel defined by \eqref{eq:momentum-kernel}.



\section{Permutation symmetric numerators  in pion parametrization} \label{sec: pion-numerators}

In the previous sections we demonstrated how to solve the NLSM kinematic
numerators via the KLT inspired approach using Cayley parametrized
currents as inputs. We obtained $(n-2)!$ half ladder basis numerators with specifically
chosen pairs of legs $(1,\,n)$ fixed at two ends,
\begin{eqnarray}
  \begin{minipage}{1cm}  \includegraphics[width=3cm]{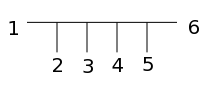} \end{minipage} \hspace{2cm} & = &  n_{1|2345|6},\\
 \begin{minipage}{1cm}   \includegraphics[width=3cm]{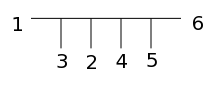} \end{minipage}  \hspace{2cm} &  = &  n_{1|3245|6},\\
 & \vdots\nonumber
\end{eqnarray}
 while all other numerators were determined by these basis half ladders
through anti-symmetry and Jacobi identities. For example the following
 numerators are given by
\begin{eqnarray}
 \begin{minipage}{1cm}  \includegraphics[width=3cm]{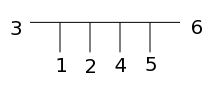}  \end{minipage} \hspace{2cm} & = & (-1)\times   \begin{minipage}{1cm}  \includegraphics[width=3cm]{n132456}  \end{minipage} \hspace{2cm}  \label{eq:6pt-antisymm}\\
 & = & (-1)\, n_{1|3245|6}\nonumber
\\
\nonumber \\
  \begin{minipage}{1cm} \includegraphics[width=3cm]{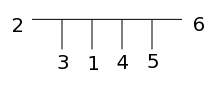}  \end{minipage} \hspace{2cm} & = & (-1)\times  \begin{minipage}{1cm}    \includegraphics[width=3cm]{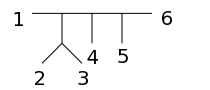} \end{minipage} \hspace{2cm} \label{eq:6pt-jacobi}\\
 & = & (-1)\times\left(  \begin{minipage}{1cm}  \includegraphics[width=3cm]{n123456}  \end{minipage} \hspace{2cm} -  \begin{minipage}{1cm}   \includegraphics[width=3cm]{n132456} \end{minipage} \hspace{2cm} \right)\nonumber \\
 & = &  (-1)\left(n_{1|2345|6}-n_{1|3245|6}\right)\nonumber
\end{eqnarray}
The Jacobi identities were satisfied trivially because of the conditions
we imposed to define non-basis numerators
\begin{eqnarray}
& &  \begin{minipage}{1cm}  \includegraphics[width=3cm]{n123456}   \end{minipage} \hspace{2cm}  +
  \begin{minipage}{1cm}  \includegraphics[width=3cm]{n231456}  \end{minipage} \hspace{2cm} +
   \begin{minipage}{1cm}  \includegraphics[width=3cm]{n312456}   \end{minipage} \hspace{2cm}  \\ \nonumber
   &  & =  n_{1|2345|6}+(-1)\left(n_{1|2345|6}-n_{1|3245|6}\right)+(-1)n_{1|3245|6}\\
 &  & = 0\nonumber
\end{eqnarray}
In this approach we never needed to use the KLT inspired formulas
for half ladder numerators with reference legs other than $(1,\,n)$. When appropriate
propagators  are included, all amplitudes in the KK
sector  $A(1\alpha_{2,n-1} n)$ are expressible as linear combinations of the
numerators, and the numerators in turn can be spanned by half ladders
 $n_{1|\beta_{2,n-1}|n}$. Note however that a numerator sharing the same topology
as half ladder and yet having legs other than $(1,\,n)$
fixed at two ends, for example the numerator appeared on the left
hand side of equation (\ref{eq:6pt-jacobi}), which in this picture
is \textit{defined} to be  $(-1)\left(n_{1|2345|6}-n_{1|3245|6}\right)$,
will generically be different from  $n_{2|3145|6}$, the result obtained from KLT
inspired formula, using legs $(2,\,n)$ as the references legs instead
of $(1,\,n)$. Especially that, as we can see by directly permuting
the first three legs of the explicit formula (\ref{6pt-numerator})
that  $n_{1|2345|6}+n_{2|3145|6}+n_{3|1245|6}$ does not give vanishing result.

This asymmetry was a result of the fact that, as was explained in
\cite{Chen:2013fya}, Cayley parametrization only satisfies KK relation
when all external boson lines are taken on-shell. As a matter of fact
it was proven in \cite{Fu:2014pya} that the KLT prescribed numerators
will automatically satisfies all the permutation symmetries had the KK relations
been respected off-shell by the amplitudes. As an alternative, in the remaining part of
this paper we will provide a more symmetric set of solutions for the
NLSM numerators, where Jacobi identities and anti-symmetries between
the numerators can be achieved through direct relabeling of the external
boson lines. The Cayley parametrization version comparing with the
later results nevertheless has the virtue of providing a much simpler prescription
for the numerators.


\subsection{Pion parametrization and the off-shell KK relations}

In order to avoid asymmetry let us return to the original pion model, where $U[\phi]=exp\{i\phi\}$.
For the purpose of discussion we would like to use the identity $\partial U^{-1}=U^{-1}\left(\partial U\right)U^{-1}$
and rewrite the chiral NLSM Lagrangian as the square of Mauerer-Cartan
Form
\begin{eqnarray}
S_{NLSM}[\phi] & = & tr\int\partial_{\mu}U^{-1}\partial^{\mu}U\nonumber \\
 & = & tr\int\left(\partial U\right)U^{-1}\left(\partial U\right)U^{-1}\label{eq:nlsm-action}
\end{eqnarray}
and noting that
\begin{equation}
\left(\partial U\right)U^{-1}=i\left\{ \left(\partial\phi\right)+\frac{(-i)}{2!}[\partial\phi,\phi]+\frac{(-i)^{2}}{3!}[[\partial\phi,\phi]\phi]+\frac{(-i)^{3}}{4!}[[[\partial\phi,\phi]\phi]\phi]+\dots\right\} \label{eq:commutators}
\end{equation}
For simplicity we have chosen the pion decay constant $\sqrt{2}/F=1$.
 Substituting the above into the NLSM
Lagrangian and expanding gives
\begin{eqnarray}
\mathcal{O}(\phi^{2}) & : & tr\,\partial\phi\partial\phi\label{eq:o3}\\
\mathcal{O}(\phi^{4}) & : & \frac{(-i)^{2}2}{4!}\,tr\,[[\partial\phi,\phi]\phi]\partial\phi\label{eq:o4}\\
\mathcal{O}(\phi^{6}) & : & \frac{(-i)^{4}2}{6!}\,tr\,[[[[\partial\phi,\phi]\phi]\phi]\phi]\partial\phi\label{eq:o6}\\
 &  & \vdots\nonumber \\
\mathcal{O}(\phi^{2n}) & : & \frac{(-i)^{2n-2}2}{(2n)!}tr\,[[[[\partial\phi,\phi]\phi]\dots\phi]\partial\phi\label{eq:o2n}
\end{eqnarray}
where we have used the identity $tr(A[BC])=tr([AB]C)$ between traces
to write every term in the Lagrangian as a successive commutator followed
by a $\partial\phi$, before collecting them. We are also ignoring
all odd number vertices because amplitudes of odd numbers of pions
scattering are known to be vanishing \cite{Kampf:2013vha}. Note from equations
(\ref{eq:o3}) to (\ref{eq:o2n}) that the color/flavor dependence
only enters the pion amplitude through a string of structure constants
\begin{equation}
f^{a_{1}a_{2}e_{1}}f^{e_{1}a_{3}e_{2}}\dots f^{e_{2n-3}a_{2n-1}a_{2n}}\times\left(\partial\phi^{a_{1}}\right)\phi^{a_{2}}\phi^{a_{3}}\dots\left(\partial\phi^{a_{2n}}\right)
\end{equation}
It is apparent that every Feynman graph in this formulation automatically
carries one copy of the color structure in the form of a cubic tree,
and the anti-symmetry with respect to swapping any two neighboring
branches in the cubic structure guarantees KK relations between pion
amplitudes even when off-shell. As we shall see in the following discussions
that pion currents constructed from the vertices above can be used
to determine symmetric BCJ numerators.

The Feynman rules in this formulation of the NLSM however does not
directly prescribe a Jacobi satisfying BCJ numerators, as can
be verified by naively identifying all the momentum dependent factors
associated with each color tree as the corresponding kinematic
numerator and then examine their Jacobi sum. This is because the color
cubic trees are not independent, and each double-copy structure is
in fact a linear combination of the Feynman graphs.

The color/flavor ordered vertices in this model are given by
\begin{eqnarray}
V_{4}(1234) & = & \frac{(-i)^{4}2}{4!}\times\biggl\{  \left(s_{12}+s_{23}+s_{34}+s_{41}\right)-2\left(s_{13}+s_{24}+s_{31}+s_{42}\right)+\left(s_{14}+s_{21}+s_{32}+s_{43}\right)\biggr\}\nn\label{eq:4-vertex}\\
V_{6}(123456) & = & \frac{(-i)^{6}2}{6!}\times\biggl\{\left(s_{12}+s_{23}+s_{34}+s_{45}+s_{56}+s_{61}\right)-4\left(s_{13}+s_{24}+s_{35}+s_{46}+s_{51}+s_{62}\right)\nn
 &  & +6\left(s_{14}+s_{25}+s_{36}+s_{41}+s_{52}+s_{63}\right)-4\left(s_{15}+s_{26}+s_{31}+s_{42}+s_{53}+s_{64}\right)\nonumber \\
 &  & +\left(s_{16}+s_{21}+s_{32}+s_{43}+s_{54}+s_{65}\right)\biggr\}\label{eq:6-vertex} \\
 & \vdots\nonumber
\end{eqnarray}
Generically a $2n$-vertex reads
\begin{eqnarray}
V_{2n}(12\dots,2n) & = \frac{(-i)^{4}2}{(2n)!}\times  \biggl\{  & \left(s_{12}+s_{23}+\dots\right)-\left(\begin{array}{c}
2n-2\\
1
\end{array}\right)\left(s_{13}+s_{24}+\dots\right)    \\
 & &  +\left(\begin{array}{c}
2n-2\nonumber
2
\end{array}\right)\left(s_{14}+s_{25}+\dots\right)   +\dots \nonumber \\
  & &  + \left(s_{1,2n}+s_{2,1}+\dots\right)    \biggr\}  \nonumber \\
 & = & \frac{(-i)^{2n}2}{(2n)!}\left\{ \sum_{k=1}^{2n-1}\left((-1)^{k-1}\left(\begin{array}{c}
2n-2\\
k-1
\end{array}\right)\sum_{i=1}^{2n}s_{i,i+k}\right)\right\} ,
\end{eqnarray}
which agrees with the formula derived from a slightly different approach
in \cite{Kampf:2013vha}.

As a quick check we verify the $U(1)$ decoupling identity between
$6$-Goldstone boson currents $J(12345)+J(21345)+J(23145)+J(23415)+J(23451)=0$,
which is a limited case of the more general KK relations. The $6$-vertex
contribution to the sum of amplitudes reads
\begin{eqnarray}
\left(s_{16}+s_{21}+s_{32}+s_{43}+s_{54}+s_{65}\right) & -4\left(s_{26}+s_{31}+s_{42}+s_{53}+s_{64}+s_{15}\right) & +6\left(s_{36}+s_{14}+s_{25}\right)\\
\left(s_{26}+s_{12}+s_{31}+s_{43}+s_{54}+s_{65}\right) & -4\left(s_{16}+s_{32}+s_{41}+s_{53}+s_{64}+s_{25}\right) & +6\left(s_{36}+s_{24}+s_{15}\right)\nonumber \\
\left(s_{26}+s_{32}+s_{13}+s_{41}+s_{54}+s_{65}\right) & -4\left(s_{36}+s_{12}+s_{43}+s_{51}+s_{64}+s_{25}\right) & +6\left(s_{16}+s_{24}+s_{35}\right)\nonumber \\
\left(s_{26}+s_{32}+s_{43}+s_{14}+s_{51}+s_{65}\right) & -4\left(s_{36}+s_{42}+s_{13}+s_{54}+s_{61}+s_{25}\right) & +6\left(s_{46}+s_{21}+s_{35}\right)\nonumber \\
\left(s_{26}+s_{32}+s_{43}+s_{54}+s_{15}+s_{61}\right) & -4\left(s_{36}+s_{42}+s_{53}+s_{14}+s_{65}+s_{21}\right) & +6\left(s_{46}+s_{25}+s_{31}\right)\nonumber
\end{eqnarray}
We see that every $s_{i,j}$ cancel completely without referring to
massless condition. A similar pairwise cancelation occurs among the
$4$-vertex contribution.


\subsection{BCJ relations in the pion parametrization scheme}

Following the same spirit that lead to systematic construction of
the numerators in Cayley parameterization, we would then like to apply
KLT inspired prescription on the more symmetric pion model, in the
hope that perhaps a better understanding of the algebraic structure
can be obtained in this parametrization scheme, since in this picture
the numerators are likely to be less obscured by generalized gauge
degrees of freedoms. For this purpose we need an off-shell continuation
of the BCJ relation compatible with the Feynman rules, analogous to
equation (\ref{off-shell-BCJ}). As
it turns out, off-shell BCJ relation contains richer structures in
the pion parametrization scheme: It is straightforward to verify using
the explicit formula (\ref{eq:4-vertex}) for color/flavor ordered
$4$-vertex that the BCJ sum of Goldstone boson currents at $4$-points
is given by
\begin{eqnarray}
s_{31}J(132)+(s_{31}+s_{32})J(123) & = & (s_{31}-s_{32})J(1)J(2)\times(-1)\frac{2}{4!}.\label{eq:4pt-pion-bcj}
\end{eqnarray}
We formally retain the $2$-point currents $J(i)$ on the right hand
side of the equation. This will prove helpful in recognizing the
general pattern of off-shell BCJ relations in the discussion below,
even though they are just a numerical factor one. At $6$-points we
see that a new term is being produced. In addition to products of
two sub-currents the pion parametrization scheme permits a four sub-current term,
\begin{eqnarray}
 &  & s_{51}J(15234)+(s_{51}+s_{52})J(12534)+\dots+(s_{51}+s_{52}+s_{53}+s_{54})J(12345)\label{eq:6pt-pion-bcj}\\
 &  & =(s_{51}-3s_{52}+3s_{53}-s_{54})J(1)J(2)J(3)J(4)\times(-1)^{2}\frac{2}{6!}\,\frac{1}{2}\nonumber \\
 &  & +(s_{5,1+2+4}-s_{54})J(123)J(4)\times(-1)\frac{2}{4!}\nonumber \\
 &  & +(s_{51}-s_{5,2+3+4})J(1)J(234)\times(-1)\frac{2}{4!}.\nonumber
\end{eqnarray}
Repeating a similar calculation, using the explicit formula for flavor
ordered $8$-vertex gives
\begin{eqnarray}
 &  & s_{71}J(1723456)+\dots+(s_{71}+s_{72}+\dots+s_{76})J(1234567)\label{eq:8pt-pion-bcj}\\
 &  & =(s_{71}-5s_{72}+10s_{73}-10s_{74}+5s_{75}-s_{76})J(1)J(2)J(3)J(4)J(5)J(6)\times(-1)^{3}\frac{2}{8!}\,
 \frac{2}{3}\nonumber \\
 &  & +(s_{7,1+2+3}-3s_{74}+3s_{75}-s_{76})J(123)J(4)J(5)J(6)\times(-1)^{2}\frac{2}{6!}\,\frac{1}{2}\nonumber \\
 &  & +(s_{71}-3s_{7,2+3+4}+3s_{75}-s_{76})J(1)J(234)J(5)J(6)\times(-1)^{2}\frac{2}{6!}\,\frac{1}{2}\nonumber \\
 &  & +(s_{71}-3s_{72}+3s_{7,3+4+5}-s_{76})J(1)J(2)J(345)J(6)\times(-1)^{2}\frac{2}{6!}\,\frac{1}{2}\nonumber \\
 &  & +(s_{71}-3s_{72}+3s_{73}-s_{7,4+5+6})J(1)J(2)J(3)J(456)\times(-1)^{2}\frac{2}{6!}\,\frac{1}{2}\nonumber\\
 &  & +(s_{7,1+2+3+4+5}-s_{76})J(12345)J(6)\times(-1)\frac{2}{4!}\nonumber \\
 &  & +(s_{7,1+2+3}-s_{7,4+5+6})J(123)J(456)\times(-1)\frac{2}{4!}\nonumber\\
 &  & +(s_{71}-s_{7,2+3+4+5+6})J(1)J(23456)\times(-1)\frac{2}{4!}.\nonumber
\end{eqnarray}
The pattern generalizes as we move to higher points, every time a
new term
\begin{equation}
\Biggl[\sum_{k=1}^{2n-2}(-1)^{k-1}\left(\begin{array}{c}
2n-3\\
k-1
\end{array}\right)s_{2n-1,k}\Biggr]\,J(1)J(2)J(3)\dots J(2n-2)
\end{equation}
enters the formula, distributing external Goldstone boson lines into
$(2n-2)$ sub-currents, while all splittings appeared previously in
the lower point off-shell relations carry on.

\subsubsection*{An outline of the derivations}

In this paper we calculate NLSM numerators up to $8$ points, and
equations (\ref{eq:4pt-pion-bcj}), (\ref{eq:6pt-pion-bcj}) and (\ref{eq:8pt-pion-bcj})
together are already enough for our goal. Both the derivation details of these
equations and of higher point relations follow closely
to those devised for the Cayley parametrization in \cite{Chen:2013fya}.
So instead of elaborating, in the following we provide merely an outline of the derivation.

We note that when expanded according to Berends-Giele recursion relation,
the full BCJ sum of currents can be completely broken into several
BCJ sums of the lower point currents, plus BCJ sums of vertices, to
which lower point currents are attached. The first part (namely the
BCJ sums of lower point currents) are known if we are performing the
calculation recursively. As for vertices, let us take the BCJ sum
of $4$-vertices as an example. Consider the following combination:
\begin{eqnarray}
& & s_{31}  \begin{minipage}{1cm} \includegraphics[width=3cm]{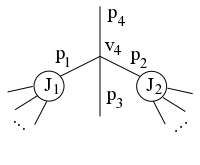}  \end{minipage} \hspace{2cm}
+(s_{31}+s_{32})  \begin{minipage}{1cm}  \includegraphics[width=3cm]{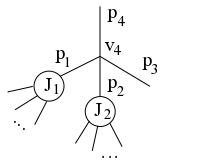}  \end{minipage} \hspace{2cm}
\\ \nonumber
& & =\left(s_{31}V_{4}(p_{1},p_{3},p_{2},p_{4})+(s_{31}+s_{32})V_{4}(p_{1},p_{2},p_{3},p_{4})\right)\,J_{1}J_{2}\frac{1}{p_{4}^{2}} \label{eq:bcj--vertex-sum}
\end{eqnarray}
Here leg $p_{3}$ is the unique leg that moves relatively to the others
in the BCJ sum and is assumed to be massless, whereas the rest of
the legs $p_{1}$, $p_{2}$ and $p_{4}$ connected to the same vertex
$V_{4}$ are not restricted to be the same. Generically $J_{1}$ and
$J_{2}$ can be sums of Feynman graphs,
\begin{eqnarray}
& &
\hspace{-4cm}  \begin{minipage}{1cm}  \includegraphics[height=2.5cm]{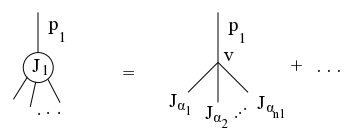}  \end{minipage} \hspace{2cm}
\label{eq:j1}  \\
& &
\hspace{-4cm}  \begin{minipage}{1cm}  \includegraphics[height=2.5cm]{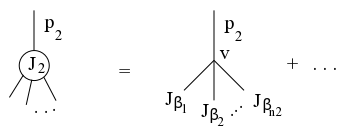}   \end{minipage} \hspace{2cm}
\label{eq:j2}
\end{eqnarray}
To proceed we note that the BCJ sum of flavor ordered $4$-vertices
can be written as the following form.

\begin{equation}
s_{31}V_{4}(p_{1},p_{3},p_{2},p_{4})+(s_{31}+s_{32})V_{4}(p_{1},p_{2},p_{3},p_{4})=p_{1}^{2}(s_{34}-s_{32})+p_{2}^{2}(s_{31}-s_{34})+p_{4}^{2}(s_{32}-s_{31})
\end{equation}
Inserting the above identity into equation (\ref{eq:bcj--vertex-sum})
produces a term proportional to $p_{1}^{2}J_{1}J_{2}\frac{1}{p_{4}^{2}}$,
a term proportional to $p_{2}^{2}J_{1}J_{2}\frac{1}{p_{4}^{2}}$,
and a term to $J_{1}J_{2}$. We then choose to let $p_{1}^{2}$ cancel
the propagator of $J_{1}$ instead of formally keeping it as $p_{1}^{2}J_{1}$.
This breaks the original $J_{1}$ into even smaller currents $V\,J_{\alpha_{1}}J_{\alpha_{2}}\dots J_{\alpha_{n1}}+\dots$.
, where the ellipsis stands for the rest of the terms in (\ref{eq:j1}).
Similarly we choose to break $p_{2}^{2}J_{2}$ into $V\,J_{\beta_{1}}J_{\beta_{2}}\dots J_{\beta_{n3}}+\dots$,
while letting the $J_{1}J_{2}$ term retain its present form. Note
that the manipulation just described is a matter of choice, however
it does affect what we get when we collect terms according to how
the full current breaks into smaller ones. Repeating the same manipulation
on BCJ sums of all vertices and collecting terms carrying the same
product of lower-point currents, and we find that all the $1/p_{4}^{2}$
pole cancel separately in each term, yielding equations (\ref{eq:4pt-pion-bcj}),
(\ref{eq:6pt-pion-bcj}) and (\ref{eq:8pt-pion-bcj}).

\subsection{Explicit numerators}

As in the case with Cayley parametrization we follow the KLT prescription
to produce NLSM numerators. At $4$ points this is simply proportional
to the BCJ sum

\begin{equation}
\sum_{\beta\in S_{2}}S[32|\beta_{2}\beta_{3}]J(1\beta_{2}\beta_{3}4)=s_{21}\Bigl(s_{31}J(132)+(s_{31}+s_{32})J(123)\Bigr),
\end{equation}
and the off-shell BCJ relation (\ref{eq:4pt-pion-bcj}) translate
the right hand side of the equation above into explicit formula
\begin{equation}
n(1234)=(-1)\frac{2}{4!}\,s_{21}(s_{31}-s_{32}).\label{eq:4pt-numerator-pion}
\end{equation}
At $6$ points we break the whole prescribed permutation sum into
first a permutation sum of leg $5$ relative to the positions of the
others, which is followed by the permutation of the rest of the legs.
The first part is again a BCJ sum, and we use (\ref{eq:6pt-pion-bcj})
to replace $6$ point currents with smaller ones, canceling a pole
$1/k_{6}^{2}$ in the process.
\begin{eqnarray}
 &  & \sum_{\beta\in S_{4}}S[5432|\beta_{2}\beta_{3}\beta_{4}\beta_{5}]J(1\beta_{2}\beta_{3}\beta_{4}\beta_{5})\\
 &  & =\sum_{\beta\in S_{3}}S[432|\beta_{2}\beta_{3}\beta_{4}]\,\Bigl(s_{51}J(15234)+(s_{51}+s_{52})J(12534)+\dots+(s_{51}+s_{52}+\dots+s_{54})J(12345)\Bigr)\nonumber\\
 &  & =\sum_{\beta\in S_{3}}S[432|\beta_{2}\beta_{3}\beta_{4}]\,(s_{51}-3s_{5\beta_{2}}+3s_{5\beta_{3}}-s_{5\beta_{4}})J(1)J(\beta_{2})J(\beta_{3})J(\beta_{4})\times\frac{2}{6!}\,\frac{1}{2}+\dots\nonumber
\end{eqnarray}
Keeping on with the same algorithm as with the Cayley parametrization
and every time canceling a pole using the off-shell BCJ relation,
yields the $6$-point NLSM numerator
\begin{eqnarray*}
 n_{1|2345|6} & = & \frac{1}{6!}\,\Biggl(-3s_{12}s_{23}\Bigl(7s_{24}s_{25}+s_{25}s_{34}+s_{15}(s_{24}+3s_{34})+s_{24}s_{35}\\
 &  & -s_{34}s_{35}+s_{14}(-s_{15}+s_{25}+3s_{35}-3s_{45})+s_{24}s_{45}-3s_{34}s_{45}\Bigr)\\
 &  & +3s_{12}s_{13}\Bigl(-s_{24}s_{25}+3s_{25}s_{34}+s_{15}(s_{24}+s_{34})+3s_{24}s_{35}\\
 &  & -s_{34}s_{35}-3s_{24}s_{45}-3s_{34}s_{45}+s_{14}(7s_{15}+s_{25}+s_{35}+s_{45})\Bigr)\Biggr)
\end{eqnarray*}
Expanding the above formula gives
\begin{eqnarray}
 n_{1|2345|6} & =\frac{3}{6!} & \Biggl(7s_{12}s_{13}s_{14}s_{15}+s_{12}s_{14}s_{15}s_{23}+s_{12}s_{13}s_{15}s_{24}-s_{12}s_{15}s_{23}s_{24}\label{eq:6pt-numerator-pion}\\
 &  & +s_{12}s_{13}s_{14}s_{25}-s_{12}s_{14}s_{23}s_{25}-s_{12}s_{13}s_{24}s_{25}-7s_{12}s_{23}s_{24}s_{25}\nonumber \\
 &  & +s_{12}s_{13}s_{15}s_{34}-3s_{12}s_{15}s_{23}s_{34}+3s_{12}s_{13}s_{25}s_{34}-s_{12}s_{23}s_{25}s_{34}\nonumber \\
 &  & +s_{12}s_{13}s_{14}s_{35}-3s_{12}s_{14}s_{23}s_{35}+3s_{12}s_{13}s_{24}s_{35}-s_{12}s_{23}s_{24}s_{35}\nonumber \\
 &  & -s_{12}s_{13}s_{34}s_{35}+s_{12}s_{23}s_{34}s_{35}+s_{12}s_{13}s_{14}s_{45}+3s_{12}s_{14}s_{23}s_{45}\nonumber \\
 &  & -3s_{12}s_{13}s_{24}s_{45}-s_{12}s_{23}s_{24}s_{45}-3s_{12}s_{13}s_{34}s_{45}+3s_{12}s_{23}s_{34}s_{45}\Biggr).\nonumber
\end{eqnarray}
As was explained in the earlier discussions we are expecting the numerators
just derived should satisfy permutation symmetries, since KK relations
are respected off-shell in the pion parametrization scheme. To check
whether this is true, first note that a $(n-2)!$ permutation symmetry
involving swapping any legs other than $(1,n)$ is a built-in feature
in the KLT prescription. Also note that leg $n$ in this prescription has
nevertheless been made special because of the off-shell continuation,
which is a price we paid to make the linear relations between amplitudes
and numerators solvable. These together leave us only the relation involving
swapping leg $1$ with any of the $(n-2)$ legs.\footnote{The fact that symmetries involving permutations of leg $n$
  is broken may not be a pathological feature of the KLT prescription. The numerator is allowed to have
  one special leg if it were to be interpreted as a string of structure constants of certain algebra,
  $n^{1,2,3,\dots}{}_{n}\sim f^{12}{}_{e_{1}}f^{e_{1}3}{}_{e_{2}}\dots f^{e_{n-3},n-1}{}_{n}$,
since we may not be neccessarily given a metric to raise the last index. } The anti-symmetry
between swapping $1\leftrightarrow2$ can be readily seen in equation
(\ref{eq:6pt-numerator-pion}). In addition we check the following
three identities.
\begin{equation}
 n_{1|2345|6}+n_{2|3145|6}+n_{3|1245|6}=0,
\end{equation}
along with
\begin{eqnarray}
 n_{4|2315|6} & = & (-1)n_{1|[[42]3]5|6}\nonumber \\
 & = & (-1)\left\{ n_{1|4235|6}-n_{1|2435|6}-n_{1|3425|6}+n_{1|3245|6}\right\} ,
\end{eqnarray}
and
\begin{eqnarray}
n_{2|3451|6} & = & (-1)n_{1|[[[23]4]5]|6}\nonumber \\
 & = & (-1)\left\{ n_{1|2345|6}-n_{1|3245|6}-n_{1|4235|6}+n_{1|4325|6}\right.\nonumber \\
 &  & \left.-n_{1|5234|6}+n_{1|5324|6}+n_{1|5423|6}-n_{1|5432|6}\right\}
\end{eqnarray}
and indeed, we find that they are all satisfied by the $6$-point
numerator (\ref{eq:6pt-numerator-pion}). We find similarly the permutation
symmetry is also satisfied by the $4$-point numerator (\ref{eq:4pt-numerator-pion}).

\subsubsection*{Rules for constructing Permutation symmetric numerators}

Following the same manipulation we obtain the $8$-point NLSM numerator
in the pion parametrization. However considering the size of the formula at $8$ points
is substantially larger  than the previous two lower point results
we shall lay out the explicit formula in appendix \ref{appendix-8pt}.
The derivations used to systematically construct permutation symmetric
numerators in this section can be summarized by the following set
of rules:
\begin{enumerate}
\item Starting with $n-2$ integers $1$, $2$, $3$, $\dots$, $n-2$ we
divide them into an even number $2m$ of ordered sets, with only an
odd number of integers allowed in every set. In addition, integer $1$ must be
assigned to the first set, and the ordering within each set does not
matter. For every such configuration we write down a factor $(\sum_{k=1}^{2m-1}(-1)^{k-1}\left(\begin{array}{c}
2m-1\\
k-1
\end{array}\right)s_{n-1,\beta_{k}})$, where $\beta_{k}$ is the sum of momenta in set $k$. For example
\begin{equation}
(134)\,(5)\,(6)\,(2)
\end{equation}
is an acceptable configuration for assigning $1$, $2$, $3$, $4$,
$5$, $6$ into four distinctive sets.  For this configuration we write
down a factor $(s_{7,1+3+4}-3s_{7,5}+3s_{7,6}-s_{7,2})$.
\item We then inspect the integers assigned into sets one by one in descending
order, if the largest integer $j$ has been assigned alone into one of the
sets, we multiply the original result by a factor $(s_{j,1}+\sum_{q\in\text{sets to the left of }j}\,\theta(j,q)s_{j,q})$,
where $\theta(j,q)=1$ if $q<j$, and $\theta(j,q)=0$ if otherwise.
To put it in plain words, we consider all integers $q$'s that are on the left
of $j$.  Whenever they are smaller than $j$ we include a factor $s_{j,q}$ into
the sum. In the example above, since integer $6$ is  assigned alone
we multiply the original result by $(s_{6,1}+s_{6,3}+s_{6,4}+s_{6,5})$.
\item Turning to the next largest integer, if it is again alone we repeat
step two; if it is not alone, we remove this integer and divide the
rest in the same set according to step one. In this example number
$4$ resides in the first set with $1$, $3$. We therefore remove integer $4$ and then divide,
producing
\begin{equation}
(1)(3)\,(5)\,(6)\,(2)
\end{equation}
Repeating the above two steps for all integers in descending order
until all integers are considered and all reside in different sets all by themselves,
the sum of products of factors we obtain following these
steps for all allowed configurations is the $n$-point numerator in
pion parametrization.
\end{enumerate}
\subsubsection*{Remarks}

We conclude this section with a few remarks. At first sight the explicit
results we presented here and in the appendix may contain an intimidatingly
large number of terms, but in fact they are much smaller than they
could be. Note that for example the $6$-point numerator has the dimension
of $(s_{i,j})^{4}$, and we have $C^{6}_{2}=15$ such Mandelstam variables, together there are $15^{4}\sim5\times10^{4}$
terms that match the dimension. Instead we only have $24$ terms
that appeared in equation (\ref{eq:6pt-numerator-pion}), all assuming
the following form
\begin{equation}
s_{i,2} \, s_{j,3} \, s_{k,4} \, s_{l,5}
\end{equation}
with $i$, $j$, $k$, $l$ only permitted to be smaller than the
label of the legs they are paired with: $l=\{1,2,3,4\}$, $k=\{1,2,3\}$, $\dots$
etc. The fact that this pattern is general can be seen from the summarizing
rules for constructing the numerators. Generically there are $(n-2)!$
terms for an $n$-point numerator $n(123\dots n)$. At the moment
it is not clear whether there is an algebraic interpretation or deeper
understanding to the structure observed, and we leave the exploration
of these questions to future works.

\section{Conclusion} \label{sec: conclusion}

From the defining Lagrangian of the theory, in this paper we have derived the BCJ
numerators of the non-linear sigma model in two different parametrization schemes.
We performed the calculation in the one leg off-shell continued scenario, where the
propagator matrix can be inverted, and the numerator expressible as the permutation
sum of currents multiplied by momentum kernel \cite{Mafra:2016ltu}. A BCJ sum was then
identified from the full expression and can be subsequently simplified using the BCJ
relation between currents, similarly to the procedure demonstrated for color-dressed
scalar theory in \cite{Du:2011js}. To proceed any step further, however, one needs to
show that the explicit form of the BCJ relation produces terms that can be again packed
into BCJ sums. And indeed, we found this is true in both the Cayley and pion parametrizations,
and this procedure was then iterated until the numerator formulas contained only Mandelstam
variables. We calculated the numerators up to $8$-points. For higher multiplicities, in both
parametrization schemes we summarize this procedure as a set of construction rules that
applies generically.  In the case of Cayley parametriztion we found that the result can
be further organized into a simple formula in terms of momentum kernel.

A perhaps most natural question brought about by these findings would then be that
whether an actual algebra can be found that explains the NLSM numerators. We note that,
provided there is enough degrees of freedom, in principle it is possible to write down an ansatz
as a string of structure constants of the most general tensorial local generators and match
the explicit numerators order by order. Alternatively, one can also obtain BCJ counterpart of
traces using for example the algorithm presented in \cite{Bern:2011ia, Du:2013sha} and try to
identify patterns. It is still not clear at the moment whether any of these approaches would
lead to simple results. It would be desirable to have a deeper understanding to the origin
of the BCJ relations observed in the NLSM.

\section*{Acknowledgments}
CF would like to thank Yannick Herfray and Kirill Krasnov for valuable discussions. CF is supported by ERC Starting Grant 277570-DIGT.
YD would like to acknowledge National Natural Science Foundation of
China under Grant Nos. 11105118, 111547310, as well as the support from 351 program of Wuhan University.

\appendix

\section{$8$-point permutation symmetric numerator}

\label{appendix-8pt}

In this appendix we list the result for $8$-point numerator in pion
parametrization.
\bea
n_{1|234567|8}=\sum_{\beta}{\cal S}[765432|\beta_{2,7}]\,J(1\beta_{2,7}8)&&
\eea
\begin{eqnarray}
=(-1)^{3} & \sum\limits_{\beta}S[65432|\beta_{2,6}] & \left(\frac{2}{8!}\right)\frac{2}{3}(s_{7,1}-5s_{7,\beta_{2}}+10s_{7,\beta_{3}}-10s_{7,\beta_{4}}+5s_{7,\beta_{5}}-s_{7,\beta_{6}})\nn
 +(-1)^{3} & \sum\limits_{\beta}S[5432|\beta_{2,5}] & \Biggl(\left(\frac{2}{4!}\right)\left(\frac{2}{6!}\right)\frac{1}{2}(s_{6,1}-s_{6,\beta_{2}})(s_{7,1+\beta_{2}+6}-3s_{7,\beta_{3}}+3s_{7,\beta_{4}}-s_{7,\beta_{5}})\nonumber \\
 &  & +\left(\frac{2}{4!}\right)\left(\frac{2}{6!}\right)\frac{1}{2}(s_{6,\beta_{2}}-s_{6,\beta_{3}})(s_{7,1}-3s_{7,6+\beta_{2}+\beta_{3}}+3s_{7,\beta_{4}}-s_{7,\beta_{5}})\nonumber \\
 &  & +\left(\frac{2}{4!}\right)\left(\frac{2}{6!}\right)\frac{1}{2}(s_{6,\beta_{3}}-s_{6,\beta_{4}})(s_{7,1}-3s_{7,\beta_{2}}+3s_{7,6+\beta_{3}+\beta_{4}}-s_{7,\beta_{5}})\nonumber \\
 &  & +\left(\frac{2}{4!}\right)\left(\frac{2}{6!}\right)\frac{1}{2}(s_{6,\beta_{4}}-s_{6,\beta_{5}})(s_{7,1}-3s_{7,\beta_{2}}+3s_{7,\beta_{3}}-s_{7,6+\beta_{4}+\beta_{5}})\nonumber \\
 &  & +\left(\frac{2}{4!}\right)\left(\frac{2}{6!}\right)\frac{1}{2}(s_{6,1}-3s_{6,\beta_{2}}+3s_{6,\beta_{3}}-s_{6,\beta_{4}})(s_{7,1+\beta_{2}+\beta_{3}+\beta_{4}}-s_{7,\beta_{5}})\nonumber \\
 &  & +\left(\frac{2}{4!}\right)\left(\frac{2}{6!}\right)\frac{1}{2}(s_{6,\beta_{2}}-3s_{6,\beta_{3}}+3s_{6,\beta_{4}}-s_{6,\beta_{5}})(s_{7,1}-s_{7,6+\beta_{2}+\beta_{3}+\beta_{4}+\beta_{5}})\Biggr)\nonumber \\
+(-1)^{3} & \sum\limits_{\beta}S[432|\beta_{2,4}] & \Biggl(\left(\frac{2}{4!}\right)\left(\frac{2}{6!}\right)\frac{1}{2}(s_{5,1}-s_{5,\beta_{2}})s_{6,1+5+\beta_{2}}(s_{7,1+5+\beta_{2}}-3s_{7,6}+3s_{7,\beta_{3}}-s_{7,\beta_{4}})\nonumber \\
 &  & +\left(\frac{2}{4!}\right)\left(\frac{2}{6!}\right)\frac{1}{2}(s_{5,1}-s_{5,\beta_{2}})s_{6,1+5+\beta_{2}+\beta_{3}}(s_{7,1+5+\beta_{2}}-3s_{7,\beta_{3}}+3s_{7,6}-s_{7,\beta_{4}})\nonumber \\
 &  & +\left(\frac{2}{4!}\right)\left(\frac{2}{6!}\right)\frac{1}{2}(s_{5,1}-s_{5,\beta_{2}})s_{6,1+5+\beta_{2}+\beta_{3}+\beta_{4}}(s_{7,1+5+\beta_{2}}-3s_{7,\beta_{3}}+3s_{7,\beta_{4}}-s_{7,6})\nonumber \\
 &  & +\left(\frac{2}{4!}\right)\left(\frac{2}{6!}\right)\frac{1}{2}(s_{5,\beta_{2}}-s_{5,\beta_{3}})s_{6,1+5+\beta_{2}+\beta_{3}}(s_{7,1}-3s_{7,5+\beta_{2}+\beta_{3}}+3s_{7,6}-s_{7,4})\nonumber \\
 &  & +\left(\frac{2}{4!}\right)\left(\frac{2}{6!}\right)\frac{1}{2}(s_{5,\beta_{2}}-s_{5,\beta_{3}})s_{6,1+5+\beta_{2}+\beta_{3}+\beta_{4}}(s_{7,1}-3s_{7,5+\beta_{2}+\beta_{3}}+3s_{7,\beta_{4}}-s_{7,6})\nonumber \\
 &  & +\left(\frac{2}{4!}\right)\left(\frac{2}{6!}\right)\frac{1}{2}(s_{5,\beta_{2}}-s_{5,\beta_{3}})s_{6,1}(s_{7,1}-3s_{7,6}+3s_{7,5+\beta_{2}+\beta_{3}}-s_{7,\beta_{4}})\nonumber \\
 &  & +\left(\frac{2}{4!}\right)\left(\frac{2}{6!}\right)\frac{1}{2}(s_{5,\beta_{3}}-s_{5,\beta_{4}})s_{6,1+\beta_{2}+\beta_{3}+\beta_{4}+5}(s_{7,1}-3s_{7,\beta_{2}}+3s_{7,5+\beta_{3}+\beta_{4}}-s_{7,6})\nonumber \\
 &  & +\left(\frac{2}{4!}\right)\left(\frac{2}{6!}\right)\frac{1}{2}(s_{5,\beta_{3}}-s_{5,\beta_{4}})s_{6,1}(s_{7,1}-3s_{7,6}+3s_{7,\beta_{2}}-s_{7,5+\beta_{3}+\beta_{4}})\nonumber \\
 &  & +\left(\frac{2}{4!}\right)\left(\frac{2}{6!}\right)\frac{1}{2}(s_{5,\beta_{3}}-s_{5,\beta_{4}})s_{6,1+\beta_{2}}(s_{7,1}-3s_{7,\beta_{2}}+3s_{7,6}-s_{7,5+\beta_{3}+\beta_{4}})\nonumber \\
 &  & +\left(\frac{2}{4!}\right)\left(\frac{2}{6!}\right)\frac{1}{2}(s_{5,1}-3s_{5,\beta_{2}}+3s_{5,\beta_{3}}-s_{s,\beta_{4}})s_{6,1+5+\beta_{2}+\beta_{3}+\beta_{4}}(s_{7,1+5+\beta_{2}+\beta_{3}+\beta_{4}}-s_{7,6})\nonumber \\
 &  & +\left(\frac{2}{4!}\right)^{3}(s_{7,1+\beta_{2}+6}-s_{7,\beta_{3}+\beta_{4}+5})(s_{6,1}-s_{6,\beta_{2}})(s_{5,\beta_{3}}-s_{5,\beta_{4}})\nonumber\\
 &  & +\left(\frac{2}{4!}\right)^{3}(s_{7,1+\beta_{2}+5}-s_{7,\beta_{3}+\beta_{4}+6})(s_{6,\beta_{3}}-s_{6,\beta_{4}})(s_{5,1}-s_{5,2})\Biggr)\nonumber
\end{eqnarray}
\begin{eqnarray*}
+(-1)^{3} &  \sum\limits_{\beta}S[32|\beta_{2,3}] & \Biggl(\left(\frac{2}{4!}\right)\left(\frac{2}{6!}\right)\frac{1}{2}(s_{4,1}-s_{4,\beta_{2}})s_{5,1+4+\beta_{2}}s_{6,1+4+\beta_{2}}(s_{7,1+4+\beta_{2}}-3s_{7,6}+3s_{7,5}-s_{7,\beta_{3}})\\
 &  & +\left(\frac{2}{4!}\right)\left(\frac{2}{6!}\right)\frac{1}{2}(s_{4,1}-s_{4,\beta_{2}})s_{5,1+4+\beta_{2}+\beta_{3}}s_{6,1+4+\beta_{2}}(s_{7,1+4+\beta_{2}}-3s_{7,6}+3s_{7,\beta_{3}}-s_{7,5})\\
 &  & +\left(\frac{2}{4!}\right)\left(\frac{2}{6!}\right)\frac{1}{2}(s_{4,1}-s_{4,\beta_{2}})s_{5,1+4+\beta_{2}}s_{6,1+4+\beta_{2}+5}(s_{7,1+4+\beta_{2}}-3s_{7,5}+3s_{7,6}-s_{7,\beta_{3}})\\
 &  & +\left(\frac{2}{4!}\right)\left(\frac{2}{6!}\right)\frac{1}{2}(s_{4,1}-s_{4,\beta_{2}})s_{5,1+4+\beta_{2}+\beta_{3}}s_{6,1+4+\beta_{2}+\beta_{3}}(s_{7,1+4+\beta_{2}}-3s_{7,\beta_{3}}+3s_{7,6}-s_{7,5})\\
 &  & +\left(\frac{2}{4!}\right)\left(\frac{2}{6!}\right)\frac{1}{2}(s_{4,1}-s_{4,\beta_{2}})s_{5,1+4+\beta_{2}}s_{6,1+4+\beta_{2}+5+\beta_{3}}(s_{7,1+4+\beta_{2}}-3s_{7,5}+3s_{7,\beta_{3}}-s_{7,6})\\
 &  & +\left(\frac{2}{4!}\right)\left(\frac{2}{6!}\right)\frac{1}{2}(s_{4,1}-s_{4,\beta_{2}})s_{5,1+4+\beta_{2}+\beta_{3}}s_{6,1+4+\beta_{2}+\beta_{3}+5}(s_{7,1+4+\beta_{2}}-3s_{7,\beta_{3}}+3s_{7,5}-s_{7,6})\\
 &  & +\left(\frac{2}{4!}\right)\left(\frac{2}{6!}\right)\frac{1}{2}(s_{4,\beta_{2}}-s_{4,\beta_{3}})s_{5,1+4+\beta_{2}+\beta_{3}}s_{6,1+4+\beta_{2}+\beta_{3}}(s_{7,1}-3s_{7,4+\beta_{2}+\beta_{3}}+3s_{7,6}-s_{7,5})\\
 &  & +\left(\frac{2}{4!}\right)\left(\frac{2}{6!}\right)\frac{1}{2}(s_{4,\beta_{2}}-s_{4,\beta_{3}})s_{5,1+4+\beta_{2}+\beta_{3}}s_{6,1+4+\beta_{2}+\beta_{3}+5}(s_{7,1}-3s_{7,4+\beta_{2}+\beta_{3}}+3s_{7,5}-s_{7,6})\\
 &  & +\left(\frac{2}{4!}\right)\left(\frac{2}{6!}\right)\frac{1}{2}(s_{4,\beta_{2}}-s_{4,\beta_{3}})s_{5,1+4+\beta_{2}+\beta_{3}}s_{6,1}(s_{7,1}-3s_{7,6}+3s_{7,4+\beta_{2}+\beta_{3}}-s_{7,5})\\
 &  & +\left(\frac{2}{4!}\right)\left(\frac{2}{6!}\right)\frac{1}{2}(s_{4,\beta_{2}}-s_{4,\beta_{3}})s_{5,1}s_{6,1+5+4+\beta_{2}+\beta_{3}}(s_{7,1}-3s_{7,5}+3s_{7,4+\beta_{2}+\beta_{3}}-s_{7,6})\\
 &  & +\left(\frac{2}{4!}\right)\left(\frac{2}{6!}\right)\frac{1}{2}(s_{4,\beta_{2}}-s_{4,\beta_{3}})s_{5,1}s_{6,1}(s_{7,1}-3s_{7,6}+3s_{7,5}-s_{7,4+\beta_{2}+\beta_{3}})\\
 &  & +\left(\frac{2}{4!}\right)\left(\frac{2}{6!}\right)\frac{1}{2}(s_{4,\beta_{2}}-s_{4,\beta_{3}})s_{5,1}s_{6,1+5}(s_{7,1}-3s_{7,5}+3s_{7,6}-s_{7,4+\beta_{2}+\beta_{3}})\\
 &  & +\left(\frac{2}{4!}\right)^{3}(s_{4,1}-s_{4,\beta_{2}})(s_{5,1+4+\beta_{2}}-s_{5,\beta_{3}})s_{6,1+5+4+\beta_{2}+\beta_{3}}(s_{7,6}-s_{7,1+5+4+\beta_{2}\beta_{3}})\\
 &  & +\left(\frac{2}{4!}\right)^{3}(s_{4,\beta_{2}}-s_{4,\beta_{3}})(s_{5,1}-s_{5,4+\beta_{2}+\beta_{3}})s_{6,1+5+4+\beta_{2}+\beta_{3}}(s_{7,6}-s_{7,1+5+4+\beta_{2}\beta_{3}})\\
 &  & +\left(\frac{2}{4!}\right)^{3}(s_{7,1+5+6}-s_{7,\beta_{2}+\beta_{3}+4})(s_{6,1}-s_{6,5})s_{5,1}(s_{4,\beta_{2}}-s_{4,\beta_{3}})\\
 &  & +\left(\frac{2}{4!}\right)^{3}(s_{7,1+\beta_{2}+4}-s_{7,5+\beta_{3}+6})(s_{6,5}-s_{6,\beta_{3}})s_{5,1+\beta_{2}+4}(s_{4,1}-s_{4,\beta_{2}})\\
 &  & +\left(\frac{2}{4!}\right)^{3}(s_{7,1+\beta_{2}+4}-s_{7,\beta_{3}+5+6})(s_{6,\beta_{3}}-s_{6,5})s_{5,1+\beta_{2}+4+\beta_{3}}(s_{4,1}-s_{4,\beta_{2}})\Biggr)
\end{eqnarray*}
\begin{eqnarray*}
+(-1)^{3} & S[2|2] & \Biggl(\left(\frac{2}{4!}\right)\left(\frac{2}{6!}\right)\frac{1}{2}(s_{3,1}-s_{3,2})s_{4,1+2+3}s_{5,1+2+3}s_{6,1+2+3}(s_{7,1+2+3}-3s_{7,6}+3s_{7,5}-s_{7,4})\\
 &  & +\left(\frac{2}{4!}\right)\left(\frac{2}{6!}\right)\frac{1}{2}(s_{3,1}-s_{3,2})s_{4,1+2+3}s_{5,1+2+3}s_{6,1+2+3}(s_{7,1+2+3}-3s_{7,6}+3s_{7,4}-s_{7,5})\\
 &  & +\left(\frac{2}{4!}\right)\left(\frac{2}{6!}\right)\frac{1}{2}(s_{3,1}-s_{3,2})s_{4,1+2+3}s_{5,1+2+3}s_{6,1+2+3+5}(s_{7,1+2+3}-3s_{7,5}+3s_{7,6}-s_{7,4})\\
 &  & +\left(\frac{2}{4!}\right)\left(\frac{2}{6!}\right)\frac{1}{2}(s_{3,1}-s_{3,2})s_{4,1+2+3}s_{5,1+2+3+4}s_{6,1+2+3+4}(s_{7,1+2+3}-3s_{7,4}+3s_{7,6}-s_{7,5})\\
 &  & +\left(\frac{2}{4!}\right)\left(\frac{2}{6!}\right)\frac{1}{2}(s_{3,1}-s_{3,2})s_{4,1+2+3}s_{5,1+2+3}s_{6,1+2+3+5+4}(s_{7,1+2+3}-3s_{7,5}+3s_{7,4}-s_{7,6})\\
 &  & +\left(\frac{2}{4!}\right)\left(\frac{2}{6!}\right)\frac{1}{2}(s_{3,1}-s_{3,2})s_{4,1+2+3}s_{5,1+2+3+4}s_{6,1+2+3+4+5}(s_{7,1+2+3}-3s_{7,4}+3s_{7,5}-s_{7,6})\\
 &  & +\left(\frac{2}{4!}\right)^{3}(s_{3,1}-s_{3,2})s_{4,1+2+3}(s_{5,1+2+3}-s_{5,4})s_{6,1+5+2+3+4}(s_{7,1+5+2+3+4}-s_{7,6})\\
 &  & +\left(\frac{2}{4!}\right)^{3}(s_{7,1+2+3}-s_{7,5+4+6})(s_{6,5}-s_{6,4})s_{5,1+2+3}s_{4,1+2+3}(s_{3,1}-s_{3,2})\\
 &  & +\left(\frac{2}{4!}\right)^{3}(s_{7,1+2+3}-s_{7,4+5+6})(s_{6,4}-s_{6,5})s_{5,1+2+3+4}s_{4,1+2+3}(s_{3,1}-s_{3,2})\Biggr)
\end{eqnarray*}


\end{document}